\documentclass[pdflatex,sn-mathphys-num, iicol]{sn-jnl}
\usepackage{bm}
\usepackage{graphicx}%
\usepackage{multirow}%
\usepackage{amsmath,amssymb,amsfonts}%
\usepackage{amsthm}%
\usepackage{mathrsfs}%
\usepackage[title]{appendix}%
\usepackage{xcolor}%
\usepackage{textcomp}%
\usepackage{manyfoot}%
\usepackage{booktabs}%
\usepackage{algorithm}%
\usepackage{algorithmicx}%
\usepackage{algpseudocode}%
\usepackage{listings}%

\theoremstyle{thmstyleone}%
\theoremstyle{thmstyletwo}%

\theoremstyle{thmstylethree}%

\begin{document}

\title[Article Title]{Physics-informed neural network for model-based global computation of the asymptotic phase of limit-cycle oscillators}


\author*[1]{\fnm{Petar} \sur{Mircheski}}\email{mircheski.p.d1e6@m.isct.ac.jp}

\author[1,2]{\fnm{Hiroya} \sur{Nakao}}\email{nakao@sc.e.titech.ac.jp}

\affil*[1]{\orgdiv{Department of Systems and Control Engineering}, \orgname{Institute of Science Tokyo}, \orgaddress{\street{2-12-1 Ookayama}, \city{Meguro}, \postcode{152-8552}, \state{Tokyo}, \country{Japan}}}

\affil[2]{\orgdiv{Research Center for Autonomous Systems Materialogy}, \orgname{Institute of Science Tokyo}, \orgaddress{\street{4259 Nagatsuta}, \city{Yokohama}, \postcode{226-8501}, \state{Kanagawa}, \country{Japan}}}


\abstract{
  The asymptotic phase is the central object of phase reduction theory for limit-cycle oscillators.
  It is a scalar coordinate defined throughout the basin of attraction of a limit cycle that advances at a constant natural frequency along every trajectory.
  Computing the phase from a given mathematical model requires solving a first-order partial differential equation (PDE) over the entire basin, which is challenging due to high computation costs and complex boundary conditions.
  In this study, we propose a physics-informed neural network to efficiently calculate the asymptotic phase of limit-cycle oscillators.
  By incorporating three types of loss functions, we train a physics-informed neural network (PINN) to satisfy the governing PDE for the asymptotic phase and the adjoint equation for the phase gradient.
  The trained network delivers the phase function accurately and, via automatic differentiation,  allows the calculation of the phase gradient governing the oscillator's response to perturbations, which enables rapid entrainment closed-loop control using strong inputs.
  We validate the approach using three types of limit-cycle oscillators and demonstrate that the PINN can facilitate fast and global phase-locking control.
}




\maketitle

\section{Introduction}\label{sec:intro}

Self-sustained rhythmic oscillations are ubiquitous in nature and engineering: they appear in neural circuits~\cite{stankovski2012inference,stankovski2017coupling}, cardiac dynamics~\cite{kralemann2014reconstructing}, chemical reactions~\cite{kuramoto1984chemical, kuramoto2019concept}, power grids~\cite{rohden2012self}, and fluid flows~\cite{taira2018phase}.
These phenomena are modeled as limit-cycle oscillators in nonlinear dynamical systems theory, and phase reduction~\cite{winfree1980geometry, kuramoto1984chemical, nakao2016phase, monga2019phase, ermentrout2019recent} provides the primary analytical framework for understanding their synchronization dynamics.
\par
In phase reduction, the multidimensional oscillator state ${\bm X}$ is compressed into a single scalar variable, the asymptotic phase $\Theta({\bm X})$, defined throughout the basin of attraction so that  $d\Theta({\bm X})/dt = \omega$ holds along every trajectory, where  $\omega$ is the natural frequency of the limit cycle.
States sharing the same phase value lie on a common isochron and asymptotically converge to a single point evolving on the limit cycle~\cite{winfree1980geometry, guckenheimer1975isochrons}.
The gradient of the asymptotic phase on the limit cycle, called the phase sensitivity function (PSF)~\cite{kuramoto1984chemical} or infinitesimal phase resetting curve (iPRC)~\cite{ermentrout2010mathematical}, quantifies the linear response of the phase to weak perturbations and is the central quantity for analyzing and designing interactions between oscillators~\cite{ermentrout2010mathematical,nakao2016phase,shirasaka2017phase,wilson2022adaptive,taira2018phase}.
When the governing ODEs are known, the PSF can be computed on the limit cycle by solving the adjoint equation~\cite{ermentrout1996type,ermentrout2010mathematical,brown2004phase,nakao2016phase}.
\par
Recently, global computation of the asymptotic phase function in the entire basin of the limit cycle, not only the phase and PSF on the limit cycle, is attracting attention~\cite{nakao2005synchrony,osinga2010continuation,mauroy2012use,huguet2013computation,nakao2016phase,perez2020global,hannam2022isochron}.
It provides us with a global picture of the transient dynamics in the basin.
Moreover, it allows for the evaluation of the effect of strong perturbations given to the oscillator state far from the limit cycle on the phase, in contrast to the PSF characterizing the linear phase response on the limit cycle.
\par
However, obtaining the phase function far away from the limit cycle requires high computational costs.
The traditional method is to consider a grid of initial conditions and evolve the oscillator states until they converge to the limit cycle~\cite{nakao2016phase}.
The Fourier average~\cite{mauroy2012use}, which is a Koopman-operator based method for calculating the asymptotic phase, also requires similar computational cost.
Another method is to trace the isochrons of the PDE for the asymptotic phase, which is accurate but also computationally expensive and does not yield a globally differentiable representation of the phase and its gradient.
In the data-driven setting, polynomial regression~\cite{namura2022estimating}, Gaussian process interpolation~\cite{yamamoto2025gaussian}, sparse identification~\cite{brunton2016discovering}, and phase autoencoder methods~\cite{yawata2024phase,yawata2025phase} have been proposed for the cases where the governing equations are unknown.
Methods of dynamic mode decomposition and generalizations, including deep-learning extensions~\cite{lusch2018deep,takeishi2017learning,otto2019linearly}, have also been used to calculate the asymptotic phase~\cite{schmid2010dynamic, kutz2016dynamic, williams2015data}.
However, these data-driven methods typically require a large number of trajectories for accurate evaluation.
When the vector field of the mathematical model is available, directly solving the PDE for the asymptotic phase offers a more principled route to the phase function.

Physics-informed neural networks (PINNs)~\cite{raissi2019physics, karniadakis2021physics} solve PDEs by embedding the governing equation as a soft penalty in the training loss, with boundary and initial conditions enforced as additional terms.
It has been demonstrated that PINNs can efficiently provide numerical solutions to various types of PDEs~\cite{cai2021physics}.
They are mesh-free, yield a smooth and continuously differentiable approximation, and naturally support differentiation via automatic differentiation~\cite{paszke2019pytorch}, which makes them suitable for evaluating the asymptotic phase function.
\par

In this paper, we propose a PINN specifically tailored for the asymptotic phase.
We adopt a smooth parametrization of the asymptotic phase, which enables an algebraic formula for the asymptotic phase in terms of the network outputs and their gradients.
Moreover, we introduce the adjoint residual loss as additional physical information to improve the accuracy of the estimated phase.
The method requires only the vector field, the natural frequency, and reference phase values on the limit cycle.
No trajectory data are needed.
With a relatively small number of collocation points, it can accurately reproduce the asymptotic phase function.
\par
We validate this approach on three oscillators spanning qualitatively different dynamical regimes: the Stuart-Landau oscillator (the normal form of the supercritical Hopf bifurcation, exact analytical results), the FitzHugh-Nagumo (FHN) model (neuronal spiking), and the R\"ossler system (non-planar phase structure in three dimensions).
For each oscillator, we report quantitative phase errors against the references.
Furthermore, we demonstrate a closed-loop phase-locking controller driven directly by the trained PINN.%

\section{Phase Reduction of Limit-Cycle Oscillators}\label{sec:phase}

\subsection{Limit-cycle oscillator}

In this section, we briefly review phase reduction theory and derive the adjoint equation for the gradient of the global asymptotic phase~\cite{winfree1980geometry, kuramoto1984chemical, nakao2016phase, monga2019phase, ermentrout2019recent}.
We consider an autonomous dynamical system
\begin{equation}
  \frac{d}{dt}{\bm X}(t) = {\bm F}({\bm X}(t)),
  \label{eq:ode}
\end{equation}
where ${\bm X} \in \mathbb{R}^{d}$ is the state vector and ${\bm F} : {\mathbb R}^{d} \to {\mathbb R}^{d}$ is a smooth vector field.
We assume that Eq.~\eqref{eq:ode} possesses an exponentially stable limit-cycle solution ${\bm X}_0(t)$ satisfying ${\bm X}_0(t + T) = {\bm X}_0(t)$ with a period $T$ and frequency $\omega = 2\pi / T$, and we denote the basin of attraction of this limit cycle as $B \subseteq {\mathbb R}^{d}$.
We introduce a phase $\theta$ that increases from $0$ to $2\pi$ at a constant rate $\omega$ over one period on the limit cycle, and write ${\bm \chi}(\theta) = {\bm X}_0(\theta / \omega)$ for the state on the limit cycle at phase $\theta$.
The choice of the zero phase state $\theta= 0$ (i.e., the point ${\bm X}_0(0)$) can be taken arbitrarily on the limit cycle.

\subsection{Asymptotic phase and its governing PDE}

The asymptotic phase is a function $\Theta : B \to [0, 2\pi)$ satisfying
\begin{equation}
  \frac{d}{dt}\Theta({\bm X}(t)) = \omega, \quad {\bm X}(t) \in B,
  \label{eq:phase_flow}
\end{equation}
for every solution of Eq.~\eqref{eq:ode}.
This means the phase advances at a constant rate $\omega$ everywhere in $B$, not only on the limit cycle itself.
Points with the same value of the asymptotic phase form an isochron, and they converge to a common trajectory on the limit cycle as $t \to \infty$.

From Eqs.~\eqref{eq:ode} and~\eqref{eq:phase_flow}, the governing PDE for the asymptotic phase $\Theta$ is given by:
\begin{equation}
  \nabla\Theta({\bm X}) \cdot {\bm F}({\bm X}) = \omega,
  \quad
  {\bm X} \in B,
  \label{eq:phase_pde}
\end{equation}
where $\nabla = (\partial/\partial X_1,\ \cdots,\ \partial/\partial X_d)$ is a gradient with respect to ${\bm X} = (X_1,\ \cdots,\ X_d)$ and $\cdot$ is a scalar product of two real vectors.
This is a first-order linear PDE whose characteristics are the trajectories of Eq.~\eqref{eq:ode}.
On the limit cycle, the state ${\bm \chi}(\theta)$ characterized by a phase value $\theta$ satisfies $\Theta({\bm \chi}(\theta)) = \theta$, which serves as the boundary condition for this PDE.

\subsection{Phase sensitivity function}\label{subsec:psf}

The phase sensitivity function (PSF) is the gradient of $\Theta$ restricted to the limit cycle:
\begin{equation}
  {\bm Z}(\theta) = \nabla
  \Theta({\bm X})\big|_{{\bm X} = {\bm \chi}(\theta)}.
  \label{eq:psf_def}
\end{equation}
It satisfies the normalization condition ${\bm Z}(\theta) \cdot {\bm F}({\bm \chi}(\theta)) = \omega$, which follows directly from Eqs.~\eqref{eq:phase_pde} and~\eqref{eq:psf_def}, and can be computed by solving the adjoint equation
\begin{equation}
  \omega \frac{d}{d\theta}{\bm Z}(\theta) = -J(\theta)^\top {\bm Z}(\theta),
  \label{eq:adjoint}
\end{equation}
where $J(\theta) = D{\bm F}({\bm \chi}(\theta)) \in {\mathbb R}^{d \times d}$ is the Jacobian evaluated on the limit cycle at phase $\theta$~\cite{ermentrout1996type}.
When the oscillator receives a small perturbation ${\bm p}({\bm X},t) \in {\mathbb R}^d$, the phase $\theta(t) = \Theta({\bm X}(t))$ of the oscillator approximately evolves as

\begin{equation}
  \frac{d}{dt}\theta(t) = \omega + {\bm Z}(\theta(t)) \cdot{\bm p}({\bm \chi}(\theta), t)
  \label{eq:phase_reduction}
\end{equation}
at leading order in ${\bm p}$~\cite{kuramoto1984chemical,nakao2016phase}, which is the basis for analyzing synchronization and designing phase-reduction based controllers~\cite{moehlis2006optimal,wilson2014optimal,monga2019phase,zlotnik2013optimal,kato2021optimization,takata2021fast,mircheski2023phase,fukami2024data}.

\subsection{Global adjoint equation for the phase gradient}\label{subsec:global_adjoint}

The PSF~\eqref{eq:psf_def} is restricted to the limit cycle.
It is natural to extend the definition of PSF to the full basin by treating the phase gradient as a function of ${\bm X}$:
\begin{equation}
  {\bm Z}_g({\bm X}) \equiv \nabla
  \Theta({\bm X}), \quad {\bm X} \in B,
  \label{eq:psf_global}
\end{equation}
so that ${\bm Z}_g({\bm \chi}(\theta)) = {\bm Z}(\theta)$ recovers the PSF on the limit cycle.
The governing PDE for ${\bm Z}_g$ follows from a variational argument based on the conservation of phase difference under the flow, which can be derived by modifying the argument by Brown, Moehlis, and Holmes~\cite{brown2004phase}.
\par
Consider two neighboring trajectories of Eq.~\eqref{eq:ode} starting at ${\bm X}$ and ${\bm X} + \delta{\bm X}$, where $\delta{\bm X}$ is an infinitesimal variation.
Since $d\Theta/dt = \omega$ holds along every trajectory in $B$, both trajectories advance their phase at the same constant rate, so their instantaneous phase difference is conserved:
\begin{equation}
  \frac{d}{dt}\bigl[\Theta({\bm X}(t) + \delta{\bm X}(t)) - \Theta({\bm X}(t))\bigr] = 0.
  \label{eq:phase_diff_conserved}
\end{equation}
Expanding $\Theta({\bm X}+\delta{\bm X})$ to first order in $\delta{\bm X}$ gives
$\Theta({\bm X}+\delta{\bm X}) - \Theta({\bm X}) = {\bm Z}_g({\bm X}) \cdot \delta{\bm X} + O(\|\delta{\bm X}\|^2)$,
so Eq.~\eqref{eq:phase_diff_conserved} becomes, at leading order in $\delta {\bm X}$,
\begin{equation}
  \frac{d}{dt}\bigl[{\bm Z}_g({\bm X}(t)) \cdot \delta{\bm X}(t)\bigr] = 0.
  \label{eq:Zdx_conserved}
\end{equation}
Expanding the left side with the product rule yields
\begin{equation}
  \frac{d{\bm Z}_g}{dt} \cdot \delta{\bm X}
  \;+\;
  {\bm Z}_g \cdot \frac{d \delta{\bm X}}{dt}
  = 0.
  \label{eq:product_rule_adj}
\end{equation}
The perturbed trajectory satisfies the linearized flow equation
\begin{equation}
  \frac{d \delta{\bm X}(t)}{dt} = J({\bm X}(t)) \delta{\bm X}(t),
  \label{eq:linearized}
\end{equation}
where $J({\bm X}) = D{\bm F}({\bm X}) \in {\mathbb R}^{d \times d}$ is the Jacobian of ${\bm F}$.
Substituting~\eqref{eq:linearized} into~\eqref{eq:product_rule_adj} and noting that ${\bm Z}_g \cdot J  \delta{\bm X} = (J^\top {\bm Z}_g) \cdot \delta{\bm X}$, we obtain
\begin{equation}
  \left[\frac{d{\bm Z}_g({\bm X}(t))}{dt} + J({\bm X}(t))^\top{\bm Z}_g({\bm X}(t))\right]
  \cdot
  \delta{\bm X}(t) = 0.
  \label{eq:adj_inner_product}
\end{equation}
Since Eq.~\eqref{eq:adj_inner_product} must hold for every $\delta{\bm X}(t)$, the bracket must vanish.
Recognizing that
\begin{align}
\frac{d{\bm Z}_g ({\bm X})}{dt}
= \nabla{\bm Z}_g({\bm X}) \cdot \frac{d {\bm X}}{dt}
= \nabla{\bm Z}_g({\bm X}) \cdot  {\bm F}({\bm X}),
\end{align}
where $\nabla{\bm Z}_g = ( \nabla Z_{g,1},\ \cdots,\ \nabla Z_{g,d} ) \in {\mathbb R}^{d \times d}$ is a gradient matrix of the vector ${\bm Z}_g = (Z_{g,1},\ \cdots,\ Z_{g,d})$,
this condition reads
\begin{equation}
    \nabla{\bm Z}_g({\bm X}) \cdot {\bm F}({\bm X})  + J({\bm X})^\top {\bm Z}_g({\bm X}) = 0, \quad {\bm X} \in B.
    \label{eq:global_adjoint}
\end{equation}
Equation~\eqref{eq:global_adjoint} is a first-order adjoint PDE for ${\bm Z}_g$ that should be satisfied throughout the entire basin $B$.
\par
On the limit cycle ${\bm X} = {\bm X}_0(t) = {\bm \chi}(\theta = \omega t)$, the derivative reduces to $d{\bm Z}_g({\bm X}(t))/dt = d{\bm Z}(\theta(t))/dt$, so Eq.~\eqref{eq:adj_inner_product} yields
\begin{equation}
  \frac{d}{dt}{\bm Z}(t) = -J(t)^\top {\bm Z}(t),
  \label{eq:adjoint_time}
\end{equation}
which is the time-domain adjoint equation~\cite{ermentrout1996type} on the limit cycle, equivalent to Eq.~\eqref{eq:adjoint} after reparameterizing by phase $\theta = \omega t$.
Equation~\eqref{eq:global_adjoint} is therefore the basin-wide generalization of the adjoint equation for the PSF: whereas Eq.~\eqref{eq:adjoint} determines ${\bm Z}(\theta)$ only on the limit cycle, the adjoint PDE~\eqref{eq:global_adjoint} governs the full field ${\bm Z}_g({\bm X})$ throughout $B$.

The pair of equations~\eqref{eq:phase_pde} and~\eqref{eq:global_adjoint} provides a complete, overdetermined characterization of the phase and its gradient, and their simultaneous enforcement by the PINN is what motivates the adjoint residual loss introduced in Sec.~\ref{sec:method}.

\section{Physics-Informed Neural Network for the Asymptotic Phase}\label{sec:method}

\subsection{Parametrization of the asymptotic phase function}

We design a PINN to approximate the asymptotic phase function governed by Eqs.~\eqref{eq:phase_pde} and~\eqref{eq:global_adjoint}.
For a smooth and continuous representation, rather than the phase $\Theta({\bm X})$ itself, we train the PINN to output the trigonometric representation of the phase:
\begin{equation}
  \hat{u}({\bm X}) = \cos\Theta({\bm X}),
  \qquad
  \hat{v}({\bm X}) = \sin\Theta({\bm X}).
  \label{eq:uv}
\end{equation}
Both $\cos\Theta$ and $\sin\Theta$ are smooth, single-valued, and bounded functions of ${\bm X}$ over the entire basin of attraction.
The phase is recovered from the trained PINN as
\begin{equation}
  \Theta({\bm X}) =
  \mathrm{arctan} \frac{\hat{v}({\bm X})}{\hat{u}({\bm X})}.
  \label{eq:phase_recovery}
\end{equation}
The same approach has been taken in past studies~\cite{namura2022estimating,namura2023designing,yamamoto2025gaussian,yawata2024phase}.
This is also natural because $e^{i \Theta({\bm X})}$, rather than $\Theta({\bm X})$ itself, is a Koopman eigenfunction of the limit cycle~\cite{mauroy2013isostables,shirasaka2017phase} (see Appendix A).

Substituting Eq.~\eqref{eq:uv} into the Eq.~\eqref{eq:phase_pde} and applying the chain rule, we obtain
\begin{align}
  {\bm F}({\bm X}) \cdot \nabla\hat{u}({\bm X}) &= -\sin\Theta({\bm X})\bigl[{\bm F}({\bm X}) \cdot \nabla\Theta({\bm X})\bigr]
  = -\omega\hat{v}({\bm X}),
  \notag\\
  {\bm F}({\bm X}) \cdot \nabla\hat{v}({\bm X}) &= \cos\Theta({\bm X})\bigl[{\bm F}({\bm X}) \cdot \nabla\Theta({\bm X})\bigr]
  = \omega\hat{u}({\bm X}),
  \notag
\end{align}
so the single scalar PDE~\eqref{eq:phase_pde} for the asymptotic phase is equivalent to the pair
\begin{align}
  {\bm F}({\bm X}) \cdot \nabla\hat{u}({\bm X}) + \omega\hat{v}({\bm X}) &= 0, \label{eq:pde_u}\\
  {\bm F}({\bm X}) \cdot \nabla\hat{v}({\bm X}) - \omega\hat{u}({\bm X}) &= 0, \label{eq:pde_v}
\end{align}
whenever $(\hat{u},\hat{v}) \neq (0,0)$.
Both equations are linear in $\hat{u}$, $\hat{v}$ and their gradients, making them straightforward to evaluate via automatic differentiation during training.

From $\hat{u} = \cos\Theta$ and $\hat{v} = \sin\Theta$ we have
\begin{equation}
  \nabla\hat{u} = -\hat{v} \nabla\Theta,
  \qquad
  \nabla\hat{v} = \hat{u} \nabla\Theta,
  \label{eq:grad_uv}
\end{equation}
and
\begin{equation}
  -\hat{v} \nabla\hat{u} + \hat{u} \nabla\hat{v}
  = (\hat{u}^2 + \hat{v}^2) \nabla\Theta = \nabla\Theta,
  \label{eq:psf_formula}
\end{equation}
where the last equality uses $\hat{u}^2 + \hat{v}^2 = 1$.
Thus, the gradient of the phase function can be calculated from $\hat{u}$ and $\hat{v}$ using this relation.
In particular, the PSF on the limit cycle is
\begin{equation}
  {\bm Z}(\theta) = \bigl(-\hat{v} \nabla\hat{u} + \hat{u} \nabla\hat{v}\bigr)\big|_{{\bm X} =  {\bm \chi}(\theta)},
  \label{eq:psf_from_net}
\end{equation}
which can be computed using automatic differentiation of the PINN to evaluate $\nabla\hat{u}$ and $\nabla\hat{v}$ at any state on the limit cycle.

\subsection{Network architecture and loss functions}

We consider a PINN, represented by $f_\theta : {\mathbb R}^d \to {\mathbb R}^2$, which maps the state ${\bm X}$ to the trigonometric representation of the phase as ${\bm X} \mapsto (\hat{u}, \hat{v})$.
It consists of five fully-connected hidden layers, each with $128$ neurons and $\tanh$ activations, followed by a linear output layer.
The $\tanh$ activation is smooth and bounded, which benefits both the gradient computation in the
loss function and the regularity of the learned solution~\cite{goodfellow2016deep}.

The total training loss of the proposed PINN combines three penalty terms:
\begin{equation}
  L = w_{\rm pde} L_{\rm pde} + w_{\rm bc} L_{\rm bc} + w_{\rm adj} L_{\rm adj},
  \label{eq:total_loss}
\end{equation}
where $w_{\rm pde}$, $w_{\rm bc}$, and $w_{\rm adj}$ are weight parameters.
The three penalties on the right-hand side are defined as follows.

\subsubsection{Phase residual loss}

This loss function enforces that Eq.~\eqref{eq:phase_pde}, i.e., the PDE governing the asymptotic phase, is accurately satisfied.
The residuals of Eqs.~\eqref{eq:pde_u} and \eqref{eq:pde_v} are evaluated at $N_c$ collocation points $\{{\bm X}^{(k)}\}_{k=1}^{N_c}$ drawn from a uniform grid over a freely defined region enclosing the limit cycle in the basin $B$:
\begin{align}
  L_{\rm pde} = \frac{1}{N_c}\sum_{k=1}^{N_c}
  \Big[
    \left({\bm F}({\bm X}^{(k)}) \cdot \nabla\hat{u}^{(k)} + \omega\hat{v}^{(k)}\right)^2
\cr
    +
    \left({\bm F}({\bm X}^{(k)}) \cdot \nabla\hat{v}^{(k)} - \omega\hat{u}^{(k)}\right)^2
  \Big].
  \label{eq:loss_pde}
\end{align}

\subsubsection{Boundary condition loss}

This loss function enforces that the identity $\Theta({\bm \chi}(\theta)) = \theta$ on the limit cycle, which gives the boundary condition for the PDE~\eqref{eq:phase_pde}, is accurately satisfied.
The residuals are evaluated on the $N_b$ uniformly spaced collocation points $\{ {\bm \chi}(\theta_j) \}_{j=1}^{N_b}$ on the limit cycle, where $\theta_j = 2 \pi j / N_b$:
\begin{equation}
  L_{\rm bc} = \frac{1}{N_b}\sum_{j=1}^{N_b}
  \left[
    \left(\hat{u}({\bm \chi}(\theta_j)) - \cos\theta_j\right)^2
    +
    \left(\hat{v}({\bm \chi}(\theta_j)) - \sin\theta_j\right)^2
  \right].
  \label{eq:loss_bc}
\end{equation}
The limit cycle ${\bm \chi}(\theta)$ and the period $T$ are obtained beforehand by numerically solving the oscillator dynamics by the shooting method~\cite{press2007numerical}.

\subsubsection{Adjoint residual loss}

The third penalty directly enforces the global adjoint PDE~\eqref{eq:global_adjoint} derived in Sec.~\ref{subsec:global_adjoint} to be satisfied accurately.
We define the phase residuals as
\begin{equation}
  R_u \equiv {\bm F} \cdot \nabla\hat{u} + \omega\hat{v},
  \qquad
  R_v \equiv {\bm F} \cdot \nabla\hat{v} - \omega\hat{u}.
  \label{eq:residuals}
\end{equation}
Taking the gradients of $R_u$ and $R_v$ yields
\begin{equation}
  \nabla R_u = J^\top\nabla\hat{u} + H_u {\bm F} + \omega\nabla\hat{v},
  \qquad
  \nabla R_v = J^\top\nabla\hat{v} + H_v {\bm F} - \omega\nabla\hat{u},
  \label{eq:grad_residuals}
\end{equation}
where $H_u$ and $H_v$ are the Hessian matrices of $\hat{u}$ and $\hat{v}$ with respect to ${\bm X}$.
The adjoint loss penalizes the magnitudes of these gradients, evaluated at the same $N_c$ collocation points as for the phase residual loss:
\begin{equation}
  L_{\rm adj} = \frac{1}{N_c}\sum_{k=1}^{N_c}
  \Bigl[
    \bigl\|\nabla R_u^{(k)}\bigr\|^2
    +
    \bigl\|\nabla R_v^{(k)}\bigr\|^2
  \Bigr].
  \label{eq:loss_adj}
\end{equation}

To see the connection of this loss function to Eq.~\eqref{eq:global_adjoint}, we substitute $\nabla\hat{u} = -\hat{v} {\bm Z}_g$ and $\nabla\hat{v} = \hat{u} {\bm Z}_g$ with ${\bm Z}_g = \nabla\Theta$ (see Eq.~\eqref{eq:grad_uv}) into $\nabla R_u = 0$ and cancel terms using
Eq.~\eqref{eq:phase_pde}, obtaining
\begin{align}
\nabla R_u &= -\hat{v} \bigl[ J^\top{\bm Z}_g + \nabla{\bm Z}_g \cdot {\bm F} \bigr] = 0,
\end{align}
which should hold for any $\hat{v}$. This gives
\begin{align}
  \nabla{\bm Z}_g \cdot {\bm F} + J^\top{\bm Z}_g = 0,
  \label{eq:adj_from_res}
\end{align}
which is precisely Eq.~\eqref{eq:global_adjoint}.
The same result also follows from $\nabla R_v = 0$.
Hence, enforcing $\nabla R_u = \nabla R_v = 0$, given that $R_u = R_v = 0$ is equivalent to enforcing the global adjoint PDE in $B$.
\par
In practice, we compute $L_{\rm adj}$ by using automatic differentiation of the PINN twice: once to obtain $\nabla\hat{u}$, $\nabla\hat{v}$ (used in $R_u$, $R_v$), and a second time to differentiate $R_u$, $R_v$ with respect to ${\bm X}$.
The extra cost of second-order automatic differentiation is justified by the improvement in the accuracy of the phase gradients away from the limit cycle, where the phase gradient varies most rapidly, and the phase residual alone is insufficient to constrain the solution.

\section{Numerical Experiments}\label{sec:experiment}

\subsection{Training procedure}

Training of the proposed PINN is carried out in two sequential stages.
In the first stage, the Adam optimizer~\cite{kingma2014adam} runs for $5000$ epochs with the learning rate $10^{-3}$ and mini-batch size $100$. A multi-step learning rate scheduler reduces the rate by a factor of $0.3$ at epochs $2000$ and $4000$ to refine the global structure of the solution.
In the second stage, full-batch L-BFGS~\cite{liu1989limited} with a strong Wolfe line search is run for $500$ closure evaluations to sharpen the solution.
The implementation uses PyTorch~\cite{paszke2019pytorch}, where the gradients $\nabla\hat{u}$ and $\nabla\hat{v}$ are computed via the autograd-engine for automatic differentiation.
The default loss weights are $w_{\rm pde} = 1.0$, $w_{\rm bc} = 20.0$, and $w_{\rm adj} = 0.5$.
Oscillator-specific overrides of $w_{\rm adj}$, the batch size, and the number of L-BFGS closure evaluations are noted in each subsection.

\begin{figure*}[htbp]
  \centering
  \includegraphics[width=0.9\textwidth]{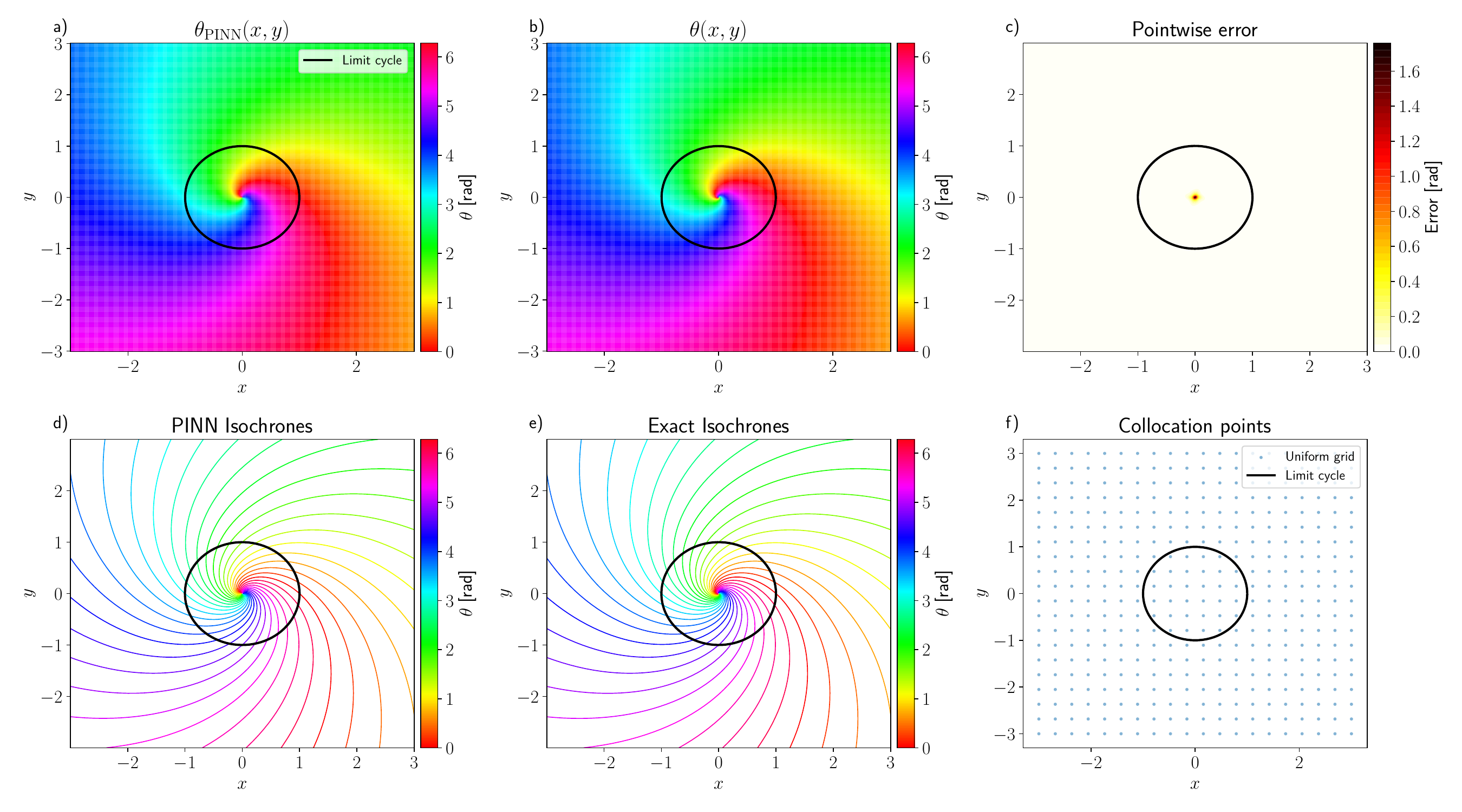}
  \caption{
    Asymptotic phase, isochrones, and data used for the Stuart-Landau oscillator.
    (a) Estimated phase function (color scale on the right) and the limit cycle (black curve).
    (b) Analytical phase function.
    (c) Pointwise error between the estimated phase function and the analytical phase function.
    (d) Estimated isochrones.
    (e) Analytically calculated isochrones.
    (f) Limit cycle and collocation points used for training.
  }
  \label{fig:sl_phase}
\end{figure*}

\begin{figure*}[t]
  \centering
  \includegraphics[width=0.9\textwidth]{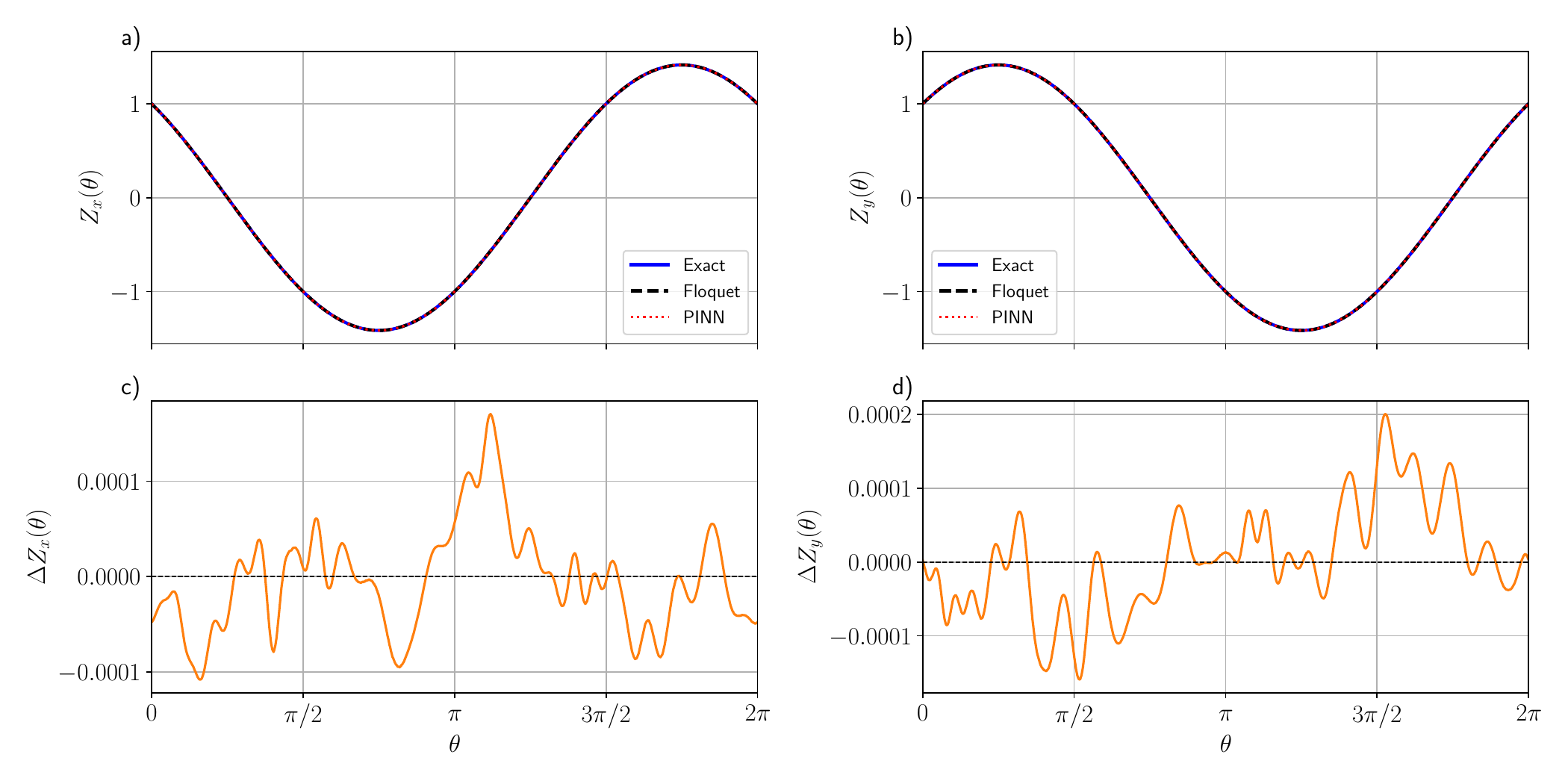}
  \caption{%
    Phase sensitivity function ${\bm Z}(\theta) = (Z_x(\theta), Z_y(\theta))$ of the Stuart-Landau oscillator.
    PINN estimate (red dots), solution of the adjoint equation~\eqref{eq:adjoint} (dashed black), and the exact analytical result (full blue line).
    (a) $Z_x(\theta) = -\sin\theta + \cos\theta$; (b) $Z_y(\theta) = \cos\theta + \sin\theta$; (c-d) Per-component difference $\Delta Z_{x,y}(\theta)$ of ${Z}_{x,y}$ between the PINN estimation and the exact solution.
  }
  \label{fig:sl_psf}
\end{figure*}

\subsection{Stuart-Landau oscillator}\label{subsec:sl}

The Stuart-Landau oscillator~\cite{kuramoto1984chemical, kuramoto2019concept}
is the normal form of a supercritical Hopf bifurcation and the simplest two-dimensional limit-cycle oscillator.
It reads
\begin{equation}
  \frac{d}{dt}
  \begin{bmatrix} x \\ y
  \end{bmatrix}
  =
  \begin{bmatrix}
    x - \alpha y - (x - \beta y)  (x^2 + y^2) \\
    \alpha x + y - (\beta x + y)  (x^2 + y^2)
  \end{bmatrix},
  \label{eq:sl}
\end{equation}
where $\alpha$ and $\beta$ are real parameters.
The limit cycle is a circle of radius $r^* = 1$ traversed at a constant angular frequency $\omega = \alpha - \beta$, so the period is $T = 2\pi/(\alpha - \beta)$.
We set $\alpha= 1.0$ and $\beta= -1.0$, giving $T = \pi$.
\par
The Stuart-Landau oscillator is unique among the systems studied here because its asymptotic phase can be exactly analytically calculated~\cite{nakao2016phase}:

\begin{equation}
  \Theta(x,y) = \arctan(y/x) - \beta \ln(\sqrt{x^2 +y^2}).
  \label{eq:sl_phase}
\end{equation}
The exact PSF on the limit cycle follows from $\nabla\Theta$,
\begin{equation}
  {\bm Z}(\theta)
  =
  \begin{pmatrix} -\sin\theta - \beta \cos\theta \\ \cos\theta - \beta\sin\theta
  \end{pmatrix},
  \label{eq:sl_psf}
\end{equation}
which is purely sinusoidal.
This analytical solution provides an exact benchmark for the PINN; any deviation from Eq.~\eqref{eq:sl_phase} or~\eqref{eq:sl_psf} directly quantifies the approximation error.

Figure~\ref{fig:sl_phase}~(f) shows the limit cycle and the collocation domain used for training.
The PINN was trained with a $20\times 20$ collocation grid covering $x,y\in[-2.0, 2.0]$, giving $400$ collocation points.
Figure~\ref{fig:sl_phase} shows the estimated phase function, the exact phase function, and a comparison against the exact reference.
The isochrons of the Stuart-Landau oscillator are logarithmic spirals: they unwind outward from the limit cycle toward the boundary of the collocation domain and converge inward toward the unstable fixed point at the origin.
\par
The PINN reproduces this spiral geometry faithfully over the entire training domain; the largest pointwise errors are concentrated near the origin, where consecutive isochrons are wound most tightly and the phase gradient grows unboundedly, making accurate phase inference inherently difficult.
Throughout the rest of the domain, including the neighborhood of the limit cycle and the outer region, the estimated phase agrees with the exact reference~\eqref{eq:sl_phase} to within a fraction of a radian.
The estimated PSF, shown in Fig.~\ref{fig:sl_psf}, almost perfectly matches the exact sinusoidal curves~\eqref{eq:sl_psf} to within plotting accuracy.

\subsection{FitzHugh-Nagumo model}\label{subsec:fhn}

The FitzHugh-Nagumo (FHN) model~\cite{fitzhugh1961impulses, nagumo1962} is a two-dimensional reduction of the Hodgkin-Huxley neuron that retains the features of neural excitability and relaxation spiking,
\begin{equation}
  \frac{d}{dt}
  \begin{bmatrix} x \\ y
  \end{bmatrix}
  =
  \begin{bmatrix} \epsilon(y + a - bx) \\ y - y^3/3 - x + I
  \end{bmatrix},
  \label{eq:fhn}
\end{equation}
where $x$ is the slow recovery variable and $y$ is the fast membrane-potential variable.
By setting the parameters as $\epsilon = 0.08$, $a = 0.7$, $b = 0.8$, $I = 0.8$, the system exhibits a stable limit cycle of period $T \simeq 36.518$ with strongly pronounced fast-slow dynamics.
The oscillation is slow most of the time, hugging one branch of the slow manifold or the other, before quickly transitioning between them through the fast manifold.
The long period and the extreme time-scale separation ($\epsilon = 0.08 \ll 1$) make this oscillator the most challenging of the three models in computing the asymptotic phase.

\begin{figure*}[htbp]
  \centering
  \includegraphics[width=\textwidth]{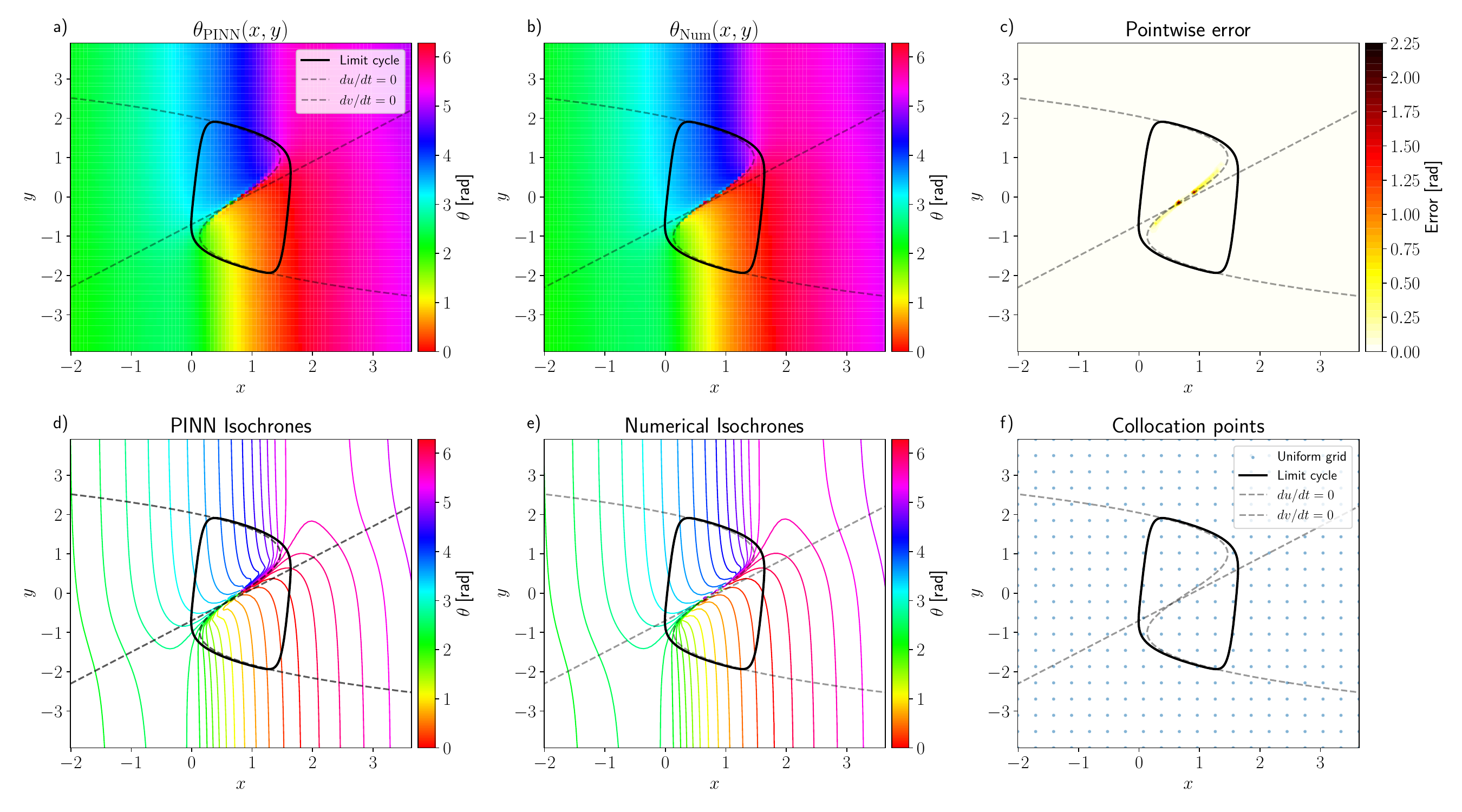}
  \caption{%
    Asymptotic phase, isochrones, and data used for the FitzHugh-Nagumo model.
    (a) Estimated phase function (color scale on the right), the limit cycle (black curve), and nullclines (gray dashed curves).
    (b) Numerically calculated phase function.
    (c) Pointwise error between the estimated phase function and the numeric phase function.
    (d) Estimated isochrones.
    (e) Numerically calculated isochrones.
    (f) Limit cycle and collocation points used.
  }
  \label{fig:fhn_phase}
\end{figure*}

\begin{figure*}[htbp]
  \centering
  \includegraphics[width=\textwidth]{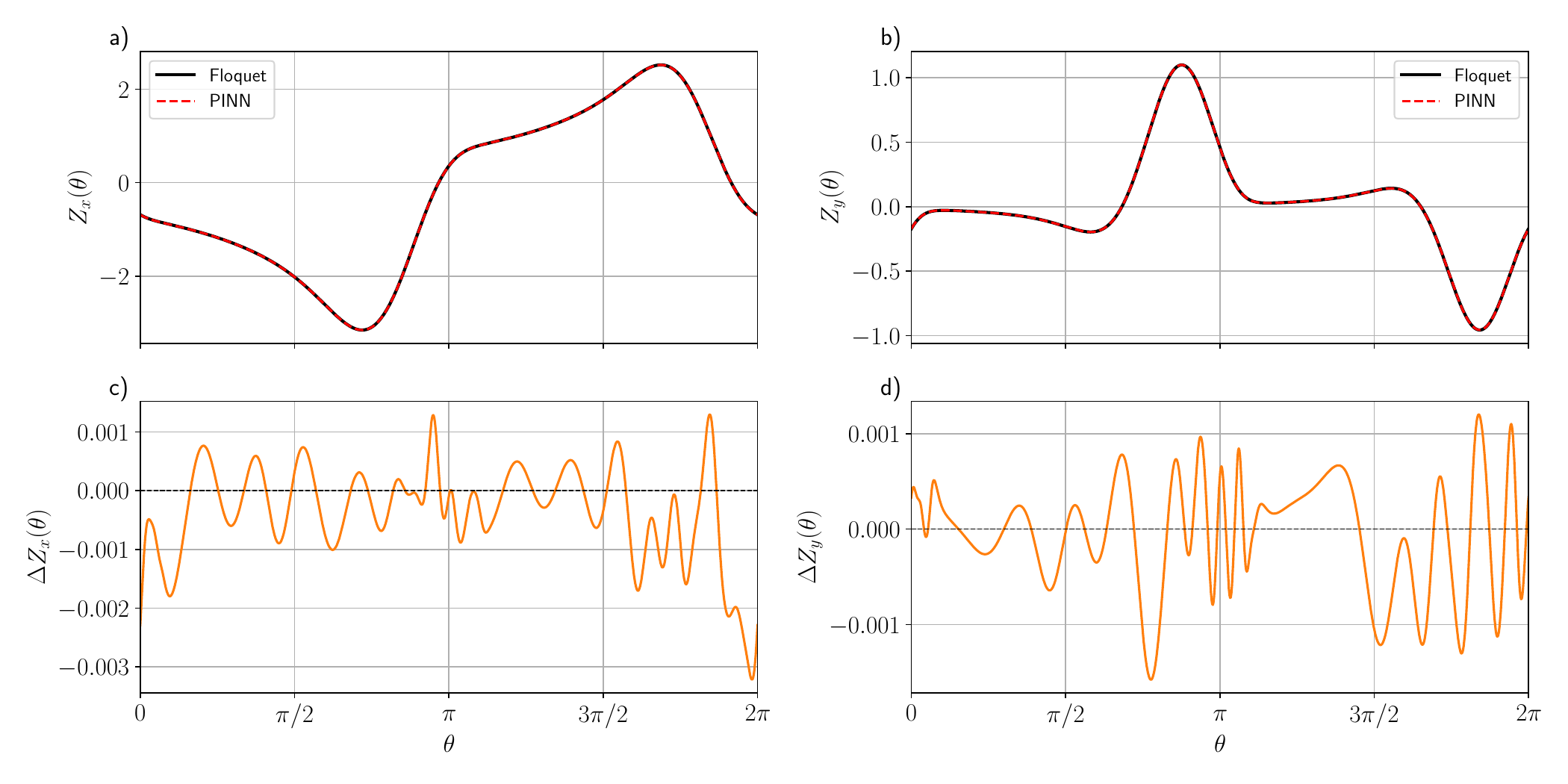}
  \caption{%
    Phase sensitivity function ${\bm Z}(\theta) = (Z_x(\theta), Z_y(\theta))$ of the FitzHugh-Nagumo oscillator.
    PINN (red dashed) vs.\ adjoint method (black solid).
    (a) $x$ component $Z_x(\theta)$; (b) $y$ component $Z_y(\theta)$; (c-d) Per-component difference $\Delta Z_{x,y}(\theta)$ between the PINN and the adjoint method.
  }
  \label{fig:fhn_psf}
\end{figure*}

Figure~\ref{fig:fhn_phase}~(f) shows the limit cycle and collocation domain.
The PINN was trained with a $20 \times 20$ collocation grid over $x \in [-2.0, 3.6]$ and $y \in [-3.9, 3.9]$.
Figure~\ref{fig:fhn_phase} shows the estimated phase and isochrons.
The most striking feature of the FHN phase portrait is the extreme compression of isochrons near the fold points of the fast-variable nullcline, the turning points at which the slow manifold loses stability, and the trajectory jumps to the opposite branch.
At these fold points, trajectories arriving from a wide range of initial conditions converge to nearly the same point before the fast transition, so many isochrons must pass through an extremely narrow strip of the state space.
This extreme bunching, which is directly tied to the small value $\epsilon = 0.08$, is the principal source of numerical difficulty: even a modest error in the PINN output translates into a large phase error in the vicinity of the nullcline.
The PINN captures the overall isochron geometry correctly and reproduces the correct qualitative structure throughout the domain;
as expected, the pointwise error is largest at the fold points and decays rapidly away from them, remaining small along the slow manifold branches where isochrons are widely spaced.
\par
The PSF, shown in Fig.~\ref{fig:fhn_psf}, has a large peak in the $y$ component (fast membrane-potential variable) at the onset of the rapid upstroke and a slower variation during the recovery phase (and also a large negative peak at the onset of the downstroke).
This structure is physiologically meaningful: it describes the brief window near the spike threshold at which the neuron's firing time is most sensitive to inputs.
The PINN reproduces the peak location and amplitude well.

\subsection{R\"ossler oscillator}\label{subsec:rossler}

The R\"ossler oscillator~\cite{rossler1976equation} is a three-dimensional autonomous system of the form
\begin{equation}
  \frac{d}{dt}
  \begin{bmatrix} x \\ y \\ z
  \end{bmatrix}
  =
  \begin{bmatrix} -y - z \\ x + a y \\ b + z (x - c)
  \end{bmatrix},
  \label{eq:rossler}
\end{equation}
with $a = 0.1$, $b = 0.1$, and $c = 4.0$, which places the system in the
period-1 limit-cycle regime (period-doubling occurs near $c \approx 6$).
The limit cycle has period $T \simeq 5.881$ and natural frequency $\omega \simeq 1.068$.
Unlike the two-dimensional oscillators above, the R\"ossler limit cycle is a three-dimensional closed orbit embedded in $\mathbb{R}^3$, and the phase function lives in a three-dimensional basin of attraction.
\par
The R\"ossler example demonstrates the straightforward extension of the PINN framework to $d = 3$: only the network input dimension is changed, while the architecture, loss functions, and training procedure remain identical.
The PINN was trained with a collocation grid embedded in a sphere with a center of the fixed point with a radius $r=6.58$ containing $8000$ points, covering the basin of attraction and the neighborhood of the limit cycle.
The adjoint regularization weight was set to $w_{\rm adj} = 5.0$, the mini-batch size was increased to $400$, and we performed $1500$ L-BFGS closure steps.

\begin{figure*}[htbp]
  \centering
  \includegraphics[width=\textwidth]{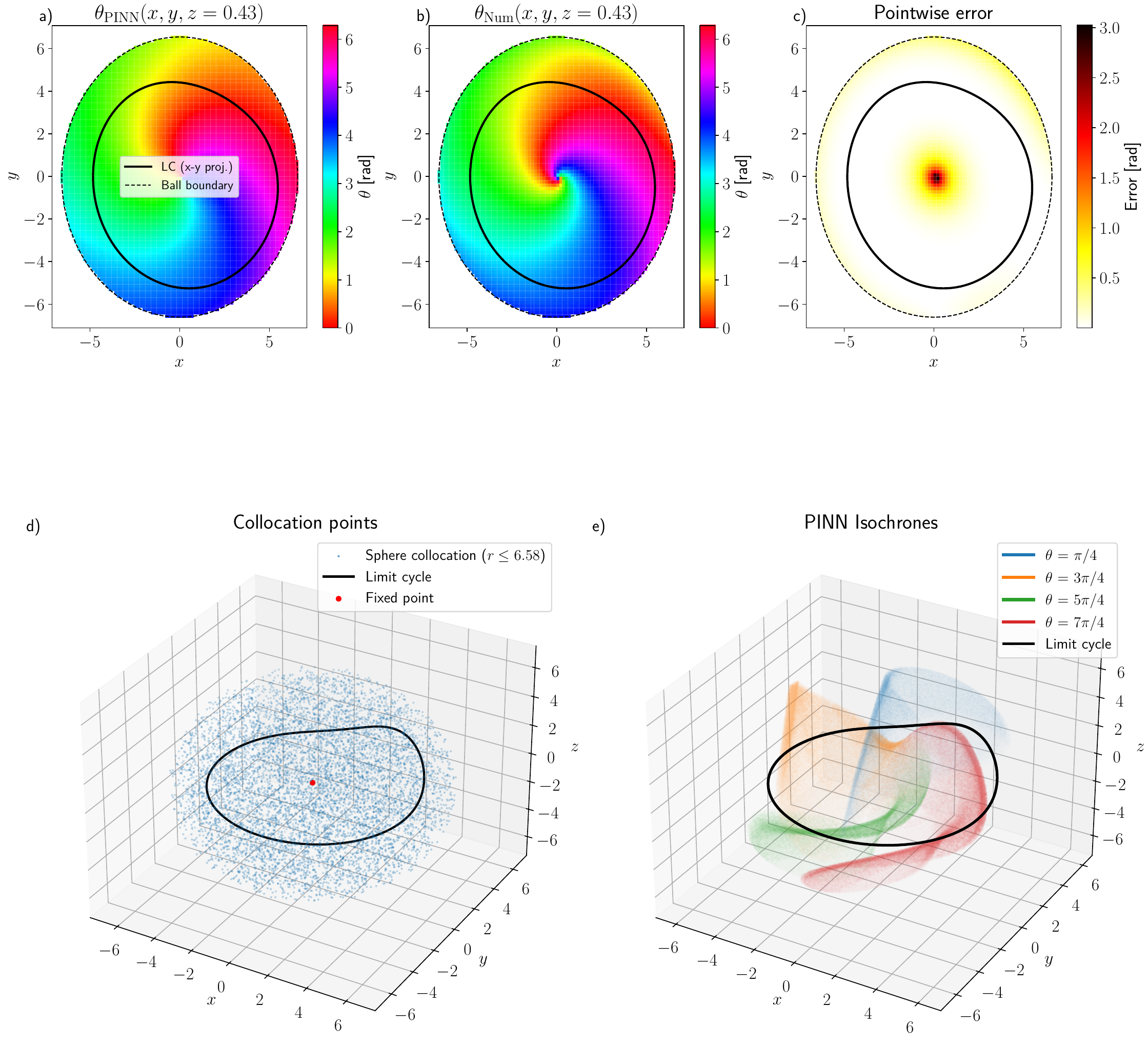}
  \caption{%
    Asymptotic phase of the R\"ossler oscillator.
    (a) Estimated phase function projected onto the $xy$-plane at $z=0.43$.
    (b) Numerically calculated phase function.
    (c) Pointwise error between the estimated phase function and the numeric phase function.
    (d) Limit cycle and collocation points used.
    (e) Four representative isochron surfaces plotted in three dimensions, colored by the phase value.
  }
  \label{fig:rossler_phase}
\end{figure*}

\begin{figure*}[htbp]
  \centering
  \includegraphics[width=0.9\textwidth]{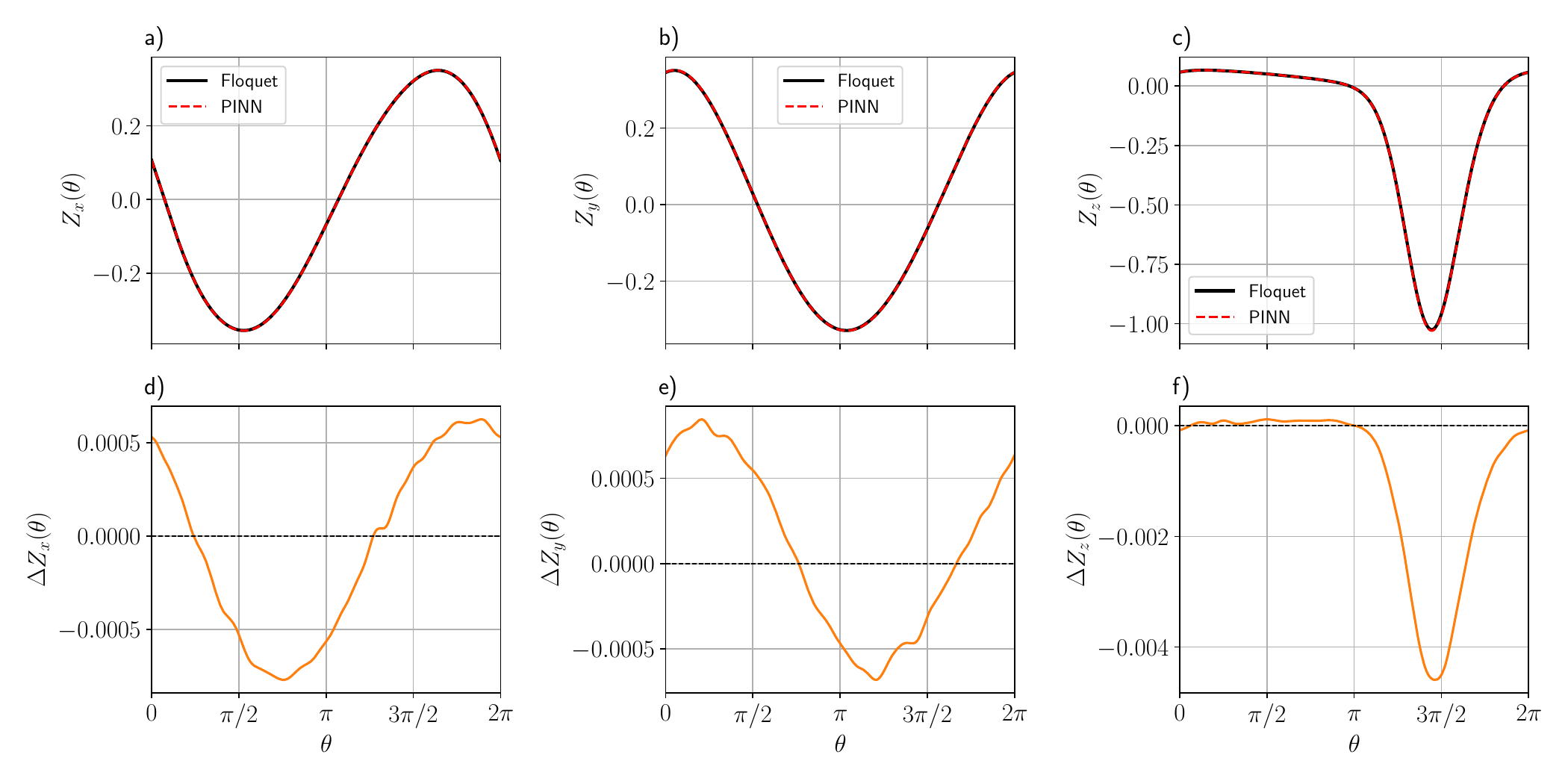}
  \caption{%
    Phase sensitivity function ${\bm Z}(\theta) = (Z_x(\theta), Z_y(\theta), Z_z(\theta))$ of the R\"ossler oscillator.
    PINN (red dashed) vs.\ adjoint method (black solid).
    (a) $x$ component $Z_x(\theta)$; (b) $y$ component $Z_y(\theta)$; (c) $Z_z(\theta)$; (d-f) Per-component difference $\Delta Z_{x,y,z}(\theta)$ between the PINN and the adjoint method.}\label{fig:rossler_psf}
\end{figure*}

Figure~\ref{fig:rossler_phase}(d) shows the three-dimensional limit cycle and collocation domain.
Figure~\ref{fig:rossler_phase} shows the phase function projected onto the $xy$-plane and a scatter comparison against the numerical reference.
The isochrons vary smoothly across the cross-section, reflecting the helical geometry of the R\"ossler orbit.
Figure~\ref{fig:rossler_phase}(e) shows four representative isochron surfaces plotted directly in three dimensions.
Each surface twists around the limit cycle, demonstrating that the PINN has learned a smooth and geometrically faithful phase function in $\mathbb{R}^3$.
Good agreement with the numerical reference is obtained throughout the collocation sphere.
The PINN recovers all three components of the PSF in good agreement with the adjoint-method result on the limit cycle.
\par
As in the planar relaxation cases, the largest pointwise errors are concentrated in the region where the limit cycle passes closest to the unstable fixed point, near the inner arc of the $xy$-projection.
In this region, the three-dimensional isochrons are most tightly packed, and the phase gradient attains its maximum magnitude, producing the same bunching that is responsible for elevated errors as in the other models.
Throughout the rest of the spherical collocation domain, where the isochrons are more broadly spaced, the estimated phase agrees with the numerical reference well.

\section{Closed-loop phase locking by a strong input}\label{sec:control}
\subsection{Control method}

As a direct application, we use the trained PINN to implement a closed-loop phase-locking controller that
entrains the oscillator's phase to a target phase $\theta^*(t) = (\omega + \Delta \omega) t = \Omega t$ rapidly by using a strong input, where $\Delta \omega$ is a frequency mismatch.
A key advantage of using the global phase function $\Theta({\bm X})$ is that the controller is valid throughout the entire basin of attraction, not merely in a small tubular neighbourhood of the limit cycle where classical phase reduction for weak perturbations holds.
\par
Consider a limit-cycle oscillator perturbed by an external input ${\bm p}(t) \in {\mathbb R}^d$:
\begin{equation}
  \frac{d}{dt}{\bm X}(t) = {\bm F}({\bm X}(t)) + {\bm p}(t).
\end{equation}
We define the instantaneous phase of this oscillator as $\theta(t) = \Theta({\bm X}(t))$.
Then, using Eq.~\eqref{eq:phase_pde}, the exact equation for $\theta$ at any state ${\bm X}$ in the basin $B$ of the limit cycle is
\begin{align}
  \frac{d}{dt}\theta(t)
   &= \nabla\Theta({\bm X}(t))\cdot{\bm F}({\bm X}(t)) + \nabla\Theta({\bm X}(t))\cdot{\bm p}(t)
   \cr
   &= \omega + {\bm Z}_g({\bm X}(t)) \cdot{\bm p}(t),
  \label{eq:global_phase_eq}
\end{align}
where ${\bm Z}_g({\bm X}) = \nabla\Theta({\bm X})$ is the spatial gradient of the phase function evaluated at the current state ${\bm X}$.
Therefore, this equation is not yet closed in $\theta(t)$ unlike the classical phase equation, but holds everywhere in the basin $B$ without approximation.
On the limit cycle, ${\bm Z}_g({\bm X})$ reduces to the PSF ${\bm Z}(\theta)$,
yielding the classical phase equation closed in $\theta$.

We choose the control input proportional to the phase gradient direction:
\begin{equation}
  {\bm p}(t) = -K {\bm Z}_g({\bm X}(t)) \sin(\Theta({\bm X}(t)) - \theta^*(t)),
  \label{eq:controller}
\end{equation}
where $K > 0$ is a gain and ${\bm Z}_g({\bm X}(t)) = \nabla\Theta({\bm X}(t))$ is computed from the trained PINN via automatic differentiation.
Substituting Eq.~\eqref{eq:controller} into Eq.~\eqref{eq:global_phase_eq} gives
\begin{equation}
  \frac{d}{dt}\theta(t) = \omega - K \|  {\bm Z}_g({\bm X}(t)) \|^2 \sin(\theta(t) - \theta^*(t)).
  \label{eq:phase_ctrl}
\end{equation}
Directing the input along ${\bm Z}_g({\bm X}) = \nabla\Theta({\bm X})$ is optimal in the sense that it maximizes the instantaneous phase-change rate per unit power of the input~\cite{monga2019phase}, and this optimality holds globally rather than only on the limit cycle.
We introduce the relative phase
\begin{equation}
\phi(t) =  \theta(t) - \theta^*(t),
\end{equation}
which is the difference between the oscillator phase and the target phase.
Then, from Eq.~\eqref{eq:phase_ctrl}, $\phi(t)$ obeys
\begin{equation}
\dot \phi(t) =  \Delta \omega - K \|  {\bm Z}_g({\bm X}(t))  \|^2 \sin \phi(t),
\label{eq:adler}
\end{equation}
which takes the form of the Adler equation but with a ${\bm X}$-dependent coefficient~\cite{adler1946}.
Because the coefficient  $K\| {\bm Z}_g(\bm{X}(t)) \|^{2}$  varies along the trajectory, Eq.~\eqref{eq:adler} is not autonomous in $\phi$ and has no true fixed point in general.
Nevertheless, as long as the inequality
\begin{equation}
  |\Delta\omega| <  K \|  {\bm Z}_g(\bm{X}(t)) \|^{2},
  \label{eq:locking_cond}
\end{equation}
is satisfied, $\phi(t)$ remains bounded and does not undergo phase slips: when $\phi$ reaches $+\pi/2$ the right-hand side of~\eqref{eq:adler}  becomes $\Delta\omega - K\|{\bm Z}_g(\bm{X}(t))\|^{2} < 0$, pushing $\phi$ back; the symmetric argument applies at $-\pi/2$.
Hence, whenever~\eqref{eq:locking_cond} holds for all $t$, the relative phase is confined to $[-\pi/2,\pi/2]$ and the oscillator stays close to the target phase without drifting away, defining the locking bandwidth of the controller.
Note that we made no assumption on the magnitude of ${\bm p}(t)$, hence the above condition can always be satisfied by setting the gain $K$ to a large enough value.
\par

\subsection{Numerical results}

\begin{figure*}[htbp]
  \centering
  \includegraphics[width=\textwidth]{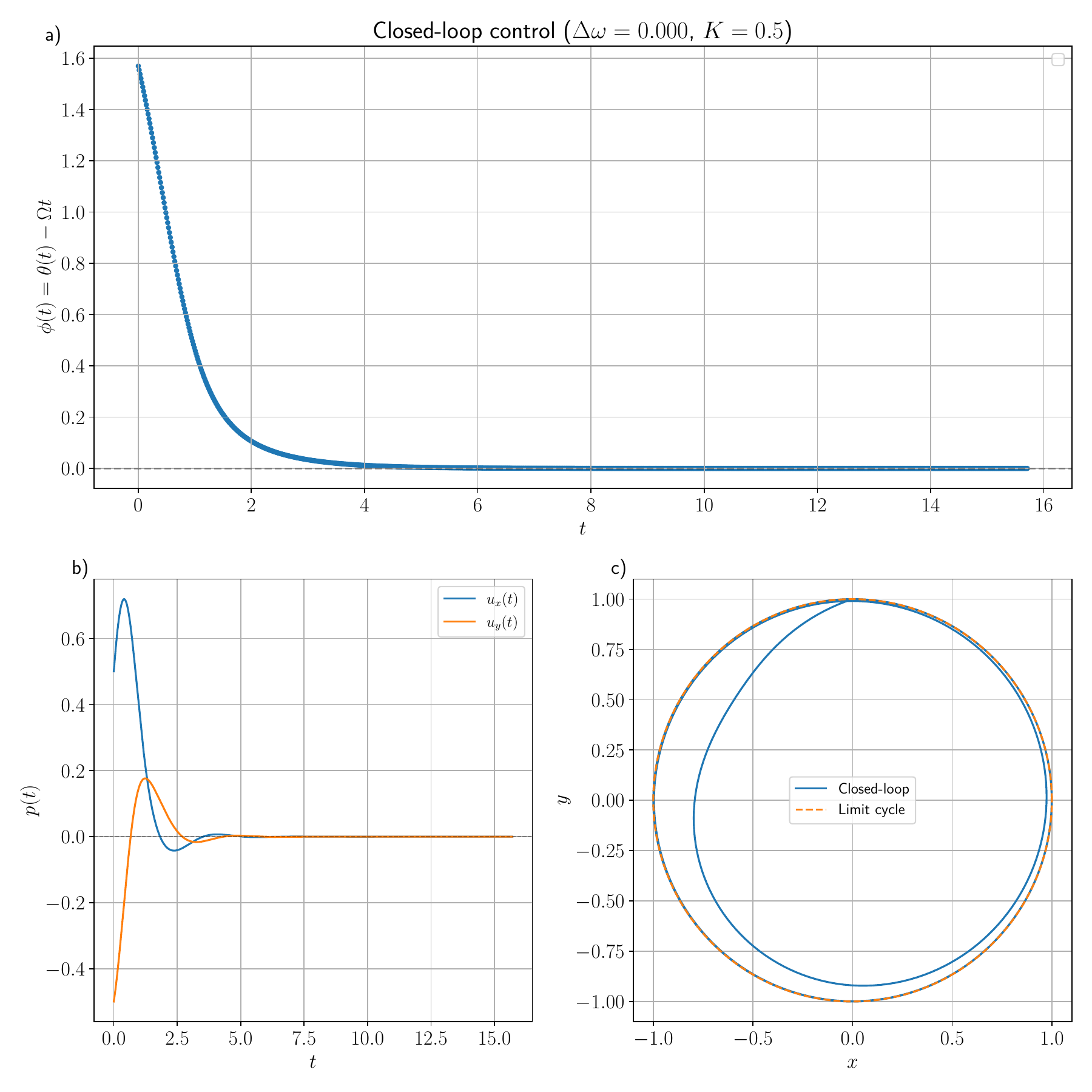}
  \caption{%
    Closed-loop phase-locking control applied to the Stuart-Landau oscillator (target phase $\theta^* = \pi$).
    (a) Phase error $\theta(t) - \theta^*$.
    (b) Control input ${\bm u}(t)$.
    (c) State-space trajectory.
  }
  \label{fig:sl_ctrl}
\end{figure*}

\begin{figure*}[htbp]
  \centering
  \includegraphics[width=0.9\textwidth]{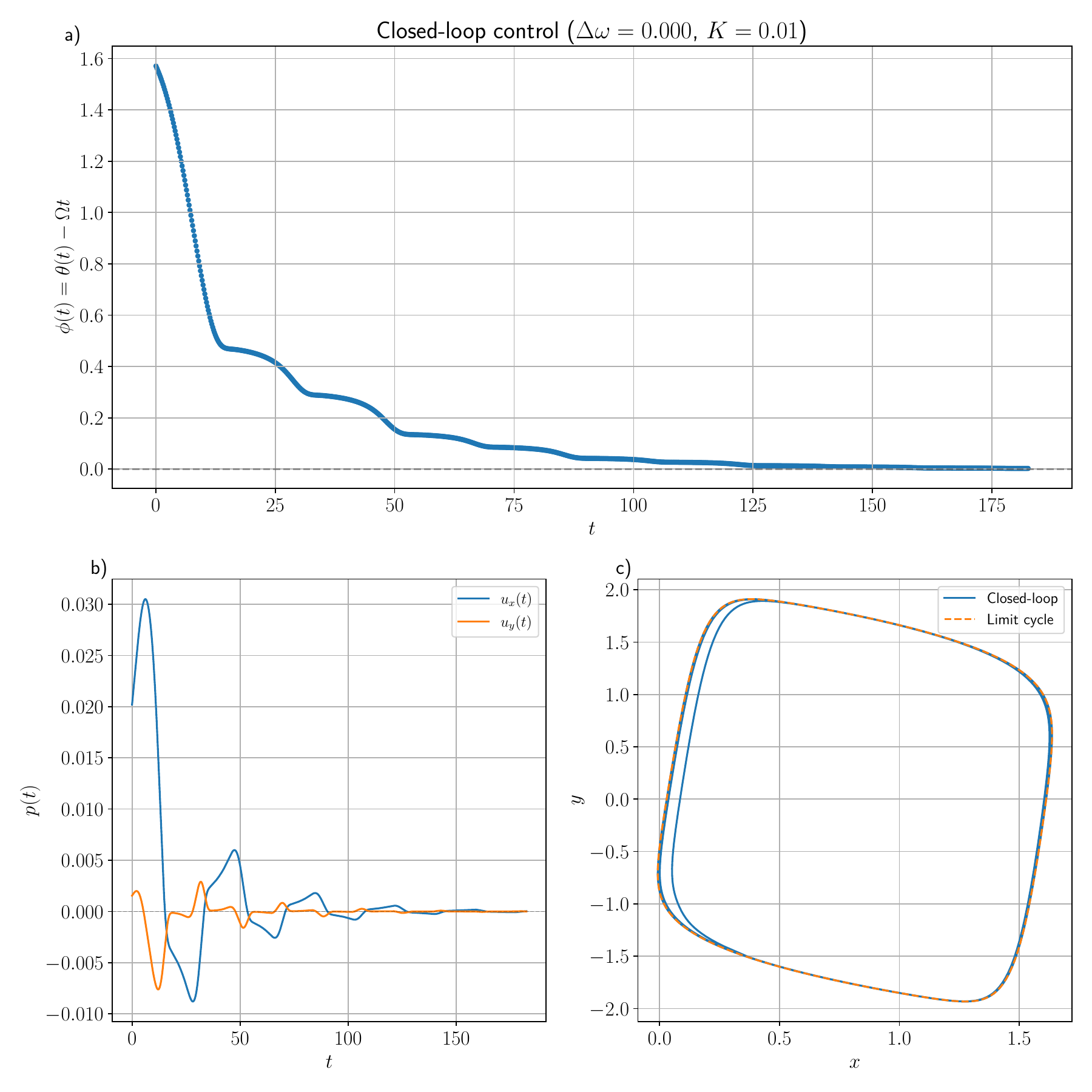}
  \caption{%
    Closed-loop phase-locking control applied to the FitzHugh-Nagumo model ($\epsilon = 0.08$, $I = 0.8$, target phase-difference $\phi^* = 0$).
    (a) Phase error $\theta(t) - \theta^*$.
    (b) Control input ${\bm u}(t)$.
    (c) State-space trajectory.
}\label{fig:fhn_ctrl}
\end{figure*}

\begin{figure*}[htbp]
  \centering
  \includegraphics[width=\textwidth]{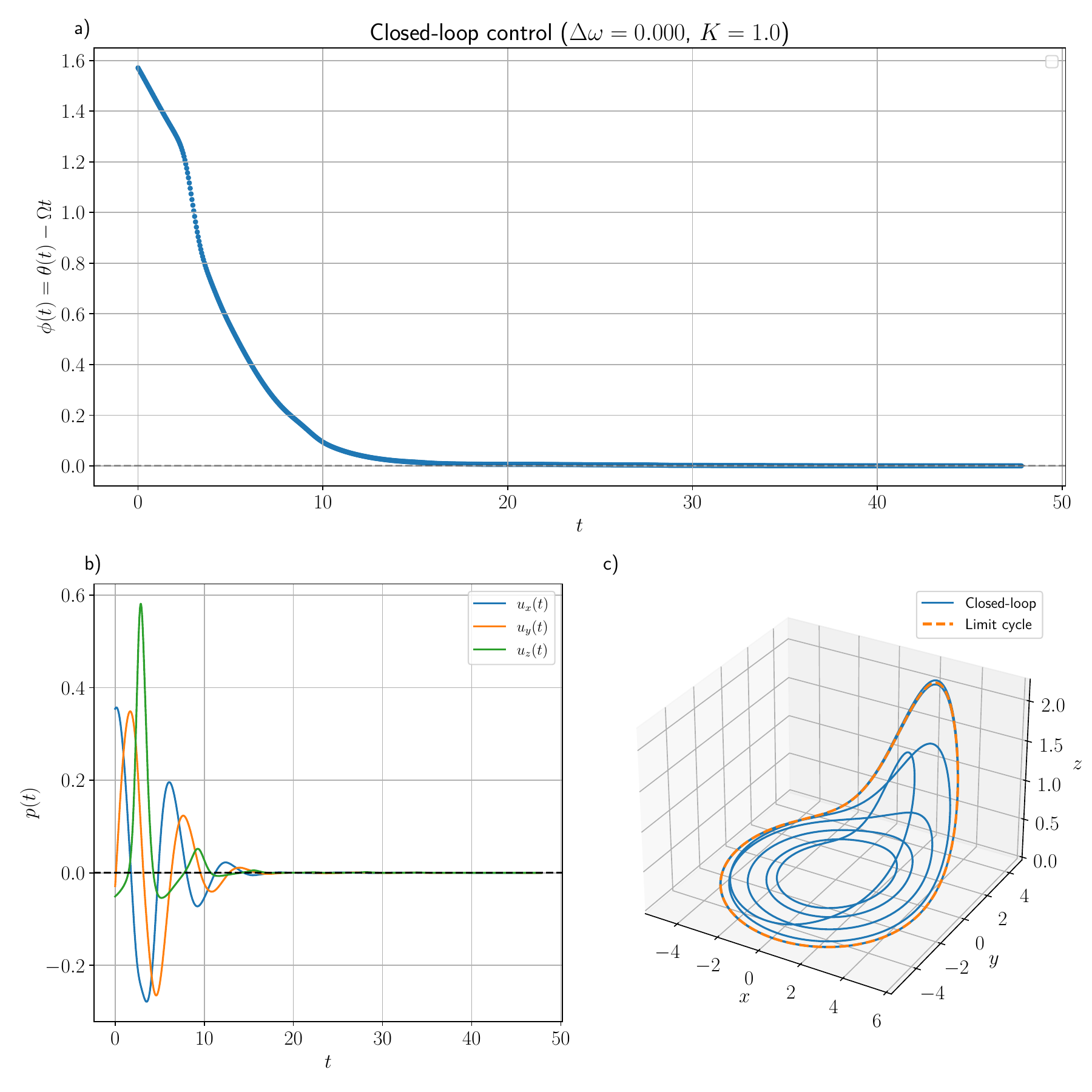}
  \caption{%
    Closed-loop phase-locking control applied to the R\"ossler oscillator ($a = 0.1$, $b = 0.1$, $c = 4.0$), target phase-difference $\phi^* = 0$.
    (a) Phase error $\theta(t) - \theta^*$.
    (b) Control input ${\bm u}(t)$.
    (c) State-space trajectory.
  }
  \label{fig:ross_ctrl}
\end{figure*}

Figures~\ref{fig:sl_ctrl}-\ref{fig:ross_ctrl} show the controller applied to all three oscillators.
In each case, the phase difference $\phi(t) = \theta(t) - \theta^*(t)$ where $\theta^*(t) = \Omega t$ starts at approximately $\pi/2$ and converges monotonically to zero, confirming that the oscillator locks to the target isochron.
\par
For the Stuart-Landau oscillator (Fig.~\ref{fig:sl_ctrl}, $K=0.5$), the relative phase decays smoothly and almost exponentially, reaching zero within roughly two natural periods ($t \lesssim 6 \approx 2T$).
The control input peaks at moderate amplitude ($|{\bm u}(t)| \lesssim 0.7$) immediately after switch-on and decays cleanly to zero without oscillation, consistent with the circular geometry of the limit cycle and the uniformly distributed phase gradient.
\par
For the FitzHugh-Nagumo model (Fig.~\ref{fig:fhn_ctrl}, $K=0.01$), the phase difference decreases rapidly during the fast-transition segments of the relaxation cycle and remains nearly flat while the trajectory moves along the slow manifold, where the phase gradient is small and the controller exerts little phase advance.
This produces the pronounced staircase structure in the convergence curve, with the control input exhibiting long flat plateaus and alternating sign reversals at the fast transitions, reflecting the strongly nonlinear, asymmetric phase sensitivity of the relaxation oscillator.
Despite the very small gain, the controller achieves locking within approximately three to four natural periods ($t \lesssim 140 \approx 4T$), and the control amplitude remains tiny throughout ($|{\bm u}(t)| \lesssim 0.03$), confirming that the phase gradient inferred by the PINN is accurate enough to steer the neuron model.
\par
For the R\"ossler oscillator (Fig.~\ref{fig:ross_ctrl}, $K=1.0$), the phase error decays smoothly, reaching zero within approximately four natural periods ($t \lesssim 25 \approx 4.5T$).
All three components of the control input oscillate briefly during the transient and then vanish, and the three-dimensional state-space trajectory spirals inward from its initial position onto the limit cycle.
The clean convergence confirms that the phase gradient $\nabla\Theta(\bm X) \in \mathbb{R}^3$ evaluated by the trained PINN is accurate enough for direct real-time feedback in the full three-dimensional state space.
\par
The consistently fast convergence across all three oscillators confirms that the global asymptotic phase function and its gradient learned by the PINN are sufficiently accurate for direct use in real-time feedback, even when the state lies far from the limit cycle.
We also note that we chose a moderate value of $K$ in each case for the purpose of demonstration. The phase equation~\eqref{eq:phase_ctrl} is valid for any value of $K$ in principle (in practice, as long as the trained PINN gives accurate phase values), because ${\bm Z}_g$ is a global quantity. Therefore, we can realize even faster entrainment by using a larger gain $K$.

\section{Discussion and Conclusions}\label{sec:conclusion}

We have proposed a physics-informed neural network for computing the global asymptotic phase function of limit-cycle oscillators.
The training loss includes three terms: the phase residual, boundary condition on the limit cycle, and adjoint residual.
In particular, the inclusion of the adjoint residual loss considerably improved the accuracy of the trained PINN.
Numerical experiments on three benchmark oscillators demonstrated the validity of the approach.
The estimated PSFs agree well with adjoint-method results in all cases, and the PINN can be used directly in a closed-loop phase-locking controller that entrains the oscillator to a target phase within a few oscillation periods.

The present method can be applied to oscillators of any dimension $d$, in principle.
The three-dimensional R\"ossler example demonstrates this directly, requiring only a change of the network's input dimension.
In practice, however, the volume of the basin grows exponentially with $d$, so higher dimensions require either more collocation points or smarter sampling strategies such as trajectory-aligned grids or adaptive importance sampling.

The present method is model-based and assumes that the vector field ${\bm F}$ is known.
If only time-series data are available, one could first identify ${\bm F}$ using sparse regression~\cite{kutz2016dynamic} or a neural ODE~\cite{chen2018neural}, and then apply the PINN to the identified model.
Alternatively, the vector field and the phase function could be learned simultaneously in a joint training scheme.
In Appendix B, a brief comparison with other methods for evaluating the asymptotic phase and phase sensitivity function are given.

A natural extension is to learn both the asymptotic phase and the isostable amplitudes~\cite{wilson2016isostable,shirasaka2017phase,mauroy2016global} within the same PINN framework by introducing additional output channels that satisfy the amplitude PDE alongside the phase PDE.
This would provide a complete phase-amplitude reduction in a single trained network.
\par
In summary, the proposed method bridges physics-informed machine learning and nonlinear oscillator theory, providing a mesh-free, globally differentiable representation of the asymptotic phase that requires only the vector field.
We expect it to be useful for oscillators whose governing equations are known but whose phase structure is difficult to characterize analytically, and as a foundation for further work on combined phase-amplitude reduction, data-free synchronization analysis, and phase-based optimal control.
\backmatter

\subsection*{Declarations}

\subsection*{Funding}
We acknowledge financial support from the JSPS KAKENHI 25H01468, 25K03081, and 22H00516

\subsection*{Competing Interests}
The authors have no relevant financial or non-financial interests to disclose

\subsection*{Data Availability}
The data that support the findings of this study are available within the article.
The source code used to generate the data and figures in this paper will be made openly available on GitHub at \url{https://github.com/PetarMirceski/pinn-phase} upon publication.

\begin{appendices}

\section{Relation to the Koopman eigenfunction}

The asymptotic phase is intimately related to the spectral theory of the Koopman operator~\cite{mauroy2013isostables,shirasaka2017phase}.
The Koopman operator $\mathcal{K}^\tau$ for the flow $\phi^\tau$ of Eq.~\eqref{eq:ode} acts on observables $g : {\mathbb R}^d \to {\mathbb C}$ by $\mathcal{K}^\tau g({\bm X}) = g(\phi^\tau({\bm X}))$.
It is a linear operator even though the underlying dynamics may be nonlinear.
The complex observable $\psi({\bm X}) = \hat{u}({\bm X}) + i\hat{v}({\bm X}) = e^{i\Theta({\bm X})}$ satisfies
\begin{equation}
  \mathcal{K}^\tau \psi = e^{i\omega\tau}\psi,
\end{equation}
i.e., $\psi$ is a Koopman eigenfunction with eigenvalue $e^{i\omega\tau}$, and the phase is its argument.
Our PINN therefore learns the principal Koopman eigenfunction of the oscillator in a model-based, PDE-constrained setting.

\section{Comparison to related methods}

Here, we briefly review existing methods for calculating the asymptotic phase in both model-based and data-driven settings.

\paragraph{Adjoint method.}

The classical adjoint method~\cite{ermentrout1996type} is a model-based method that computes the PSF accurately.
The asymptotic phase of the limit cycle can be linearly approximated by using the obtained PSF, but only in the close vicinity of the limit cycle.
We used the PSF obtained by the adjoint method to validate the accuracy of the present PINN.

\paragraph{Direct method.}

This is the most direct and fundamental method. We take a grid of initial points in the basin of the limit cycle and evolve the oscillator states until they converge to the limit cycle to evaluate their asymptotic phase values~\cite{nakao2016phase}. The computational cost can be considerably high for accurate calculation in higher dimensions.

\paragraph{Fourier average.}

This model-based method uses the fact that the asymptotic phase is the argument of the Koopman eigenfunction of the oscillator associated with the fundamental frequency, providing a simple direct formula (Fourier average) for the asymptotic phase~\cite{mauroy2012use}. This method also requires evolving the oscillator states from a grid of initial points, hence the computation cost is comparable to the direct method.

\paragraph{Isochron tracing.}

Tracing isochrons numerically as characteristics of the PDE for the asymptotic phase is an alternative model-based method for computing the asymptotic phase~\cite{osinga2010continuation,hannam2022isochron}.
It requires integrating one trajectory per phase value and per query point.
The result is smooth and accurate along the characteristics, but in the direction transversal to the characteristics, the result is a discrete table rather than a differentiable function.

\paragraph{Polynomial or Gaussian process regression.}

Polynomial regression~\cite{namura2022estimating} is the most straightforward data-driven method for parametric estimation of the asymptotic phase. Assuming that time-series data from various initial conditions are available, the phase can be estimated via polynomial fitting. However, this approach requires a substantial number of trajectories containing sufficient transient dynamics to achieve accurate results.
The Gaussian process regression~\cite{yamamoto2025gaussian} employs a kernel-based approach; it typically requires less data than polynomial regression and provides a smoother interpolation of the asymptotic phase. Nonetheless, its computational cost can become prohibitive in higher dimensions.

\paragraph{Dynamic mode decomposition (DMD).}

DMD~\cite{schmid2010dynamic} and its extensions~\cite{kutz2016dynamic, williams2015data} are data-driven methods that can estimate the Koopman eigenfunctions from time-series trajectory data.
These methods can be used to compute the asymptotic phase as the eigenfunction associated with the fundamental frequency.
Various extensions of the DMD method exist.
They provide a versatile approach, but the resulting eigenvalues and eigenfunctions may not be very accurate, because these methods simultaneously estimate multiple components by solving a projected eigenvalue problem for the data matrices, while, for the asymptotic phase, only the single eigenfunction associated with the fundamental frequency is required.

\paragraph{Koopman autoencoder and phase autoencoder.}
The Koopman autoencoder~\cite{lusch2018deep,takeishi2017learning} is a neural network approach that assumes the dynamics in the low-dimensional latent space of the autoencoder to obey the Koopman linear evolution, providing a versatile approach for dimensionality reduction and extracting low-dimensional dynamics from time-series data.
The phase autoencoder~\cite{yawata2024phase,yawata2025phase} is a more specialized neural network approach that learns the asymptotic phase from time-series data by designing a latent space in which the dynamics directly obey the phase equation, without requiring knowledge of the governing equations.
These are data-driven methods that require a sufficient amount of trajectory data, and therefore are not efficient when the mathematical model is available.
\par

For a more comprehensive overview of methods for computing the asymptotic phase and related phase reduction techniques, we refer the reader to~\cite{monga2019phase} and references therein.

In summary, the proposed PINN provides an efficient model-based approach to calculating the asymptotic phase. It does not require extensive numerical integration of the trajectories and yields accurate estimates by exploiting the high-dimensional approximation capabilities of the network.
\end{appendices}


\bibliography{sn-bibliography}

@book{winfree1980geometry,
  title           = {The Geometry of Biological Time},
  author          = {Winfree, Arthur T},
  volume          = 2,
  year            = 1980,
  doi             = {10.1007/978-3-662-22492-2},
  address         = {Berlin, Heidelberg},
  publisher       = {Springer}
}

@book{kuramoto1984chemical,
  title           = {Chemical oscillations, waves, and turbulence},
  author          = {Kuramoto, Y},
  volume          = 8,
  pages           = 156,
  year            = 1984,
  doi             = {10.1007/978-3-642-69689-3_7},
  address         = {Berlin, Heidelberg}
}

@article{stankovski2012inference,
  title           = {Inference of Time-Evolving Coupled Dynamical Systems in the
                  Presence of Noise},
  author          = {Stankovski, Tomislav and Duggento, Andrea and McClintock,
                  Peter V. E. and Stefanovska, Aneta},
  journal         = {Phys. Rev. Lett.},
  volume          = 109,
  issue           = 2,
  pages           = 024101,
  numpages        = 5,
  year            = 2012,
  month           = {Jul},
  publisher       = {American Physical Society},
  doi             = {10.1103/PhysRevLett.109.024101},
}

@article{stankovski2017coupling,
  title           = {Coupling functions: Universal insights into dynamical
                  interaction mechanisms},
  author          = {Stankovski, Tomislav and Pereira, Tiago and McClintock,
                  Peter V. E. and Stefanovska, Aneta},
  journal         = {Rev. Mod. Phys.},
  volume          = 89,
  issue           = 4,
  pages           = 045001,
  numpages        = 50,
  year            = 2017,
  month           = {Nov},
  publisher       = {American Physical Society},
  doi             = {10.1103/RevModPhys.89.045001},
}

@article{rohden2012self,
  title           = {Self-Organized Synchronization in Decentralized Power
                  Grids},
  author          = {Rohden, Martin and Sorge, Andreas and Timme, Marc and
                  Witthaut, Dirk},
  journal         = {Phys. Rev. Lett.},
  volume          = 109,
  issue           = 6,
  pages           = 064101,
  numpages        = 5,
  year            = 2012,
  month           = {Aug},
  publisher       = {American Physical Society},
  doi             = {10.1103/PhysRevLett.109.064101},
}

@article{kralemann2014reconstructing,
  title           = {Reconstructing effective phase connectivity of oscillator
                  networks from observations},
  author          = {Kralemann, Björn and Pikovsky, Arkady and Rosenblum,
                  Michael},
  journal         = {New Journal of Physics},
  year            = 2014,
  month           = {aug},
  volume          = 16,
  number          = 8,
  pages           = 085013,
  publisher       = {IOP Publishing},
  doi             = {10.1088/1367-2630/16/8/085013},
}

@article{nakao2016phase,
  author          = {Hiroya Nakao},
  title           = {Phase reduction approach to synchronisation of nonlinear
                  oscillators},
  journal         = {Contemporary Physics},
  volume          = 57,
  number          = 2,
  pages           = {188--214},
  year            = 2016,
  publisher       = {Taylor \& Francis},
  doi             = {10.1080/00107514.2015.1094987},
}

@book{ermentrout2010mathematical,
  title           = {Mathematical Foundations of Neuroscience},
  author          = {Ermentrout, Bard and Terman, David},
  volume          = 35,
  year            = 2010,
  publisher       = {Springer},
  address         = {New York, NY},
  doi             = {10.1007/978-0-387-87708-2}
}

@article{monga2019phase,
  author          = {Monga, Bharat and Wilson, Dan and Matchen, Tim and Moehlis,
                  Jeff},
  title           = {Phase reduction and phase-based optimal control for
                  biological systems: a tutorial},
  journal         = {Biological Cybernetics},
  year            = 2019,
  month           = {Apr},
  day             = 01,
  volume          = 113,
  number          = 1,
  pages           = {11-46},
  issn            = {1432-0770},
  doi             = {10.1007/s00422-018-0780-z},
}

@article{kuramoto2019concept,
  author          = {Kuramoto, Yoshiki and Nakao, Hiroya},
  title           = {On the concept of dynamical reduction: the case of coupled
                  oscillators},
  journal         = {Philosophical Transactions of the Royal Society A:
                  Mathematical, Physical and Engineering Sciences},
  volume          = 377,
  number          = 2160,
  pages           = 20190041,
  year            = 2019,
  month           = 10,
  issn            = {1364-503X},
  doi             = {10.1098/rsta.2019.0041},
}

@article{ermentrout2019recent,
  author          = {Ermentrout, Bard and Park, Youngmin and Wilson, Dan},
  title           = {Recent advances in coupled oscillator theory},
  journal         = {Philosophical Transactions of the Royal Society A:
                  Mathematical, Physical and Engineering Sciences},
  volume          = 377,
  number          = 2160,
  pages           = 20190092,
  year            = 2019,
  month           = 10,
  issn            = {1364-503X},
  doi             = {10.1098/rsta.2019.0092},
}

@article{shirasaka2017phase,
  author          = {Shirasaka, Sho and Kurebayashi, Wataru and Nakao, Hiroya},
  title           = {Phase-amplitude reduction of transient dynamics far from
                  attractors for limit-cycling systems},
  journal         = {Chaos: An Interdisciplinary Journal of Nonlinear Science},
  volume          = 27,
  number          = 2,
  pages           = 023119,
  year            = 2017,
  month           = 02,
  issn            = {1054-1500},
  doi             = {10.1063/1.4977195},
}

@article{wilson2022adaptive,
  author          = {Wilson, Dan},
  title           = {An Adaptive Phase-Amplitude Reduction Framework without
                  $\mathcal{O}(\epsilon)$ Constraints on Inputs},
  journal         = {SIAM Journal on Applied Dynamical Systems},
  volume          = 21,
  number          = 1,
  pages           = {204-230},
  year            = 2022,
  doi             = {10.1137/21M1391791},
}

@article{taira2018phase,
  title           = {Phase-response analysis of synchronization for periodic
                  flows},
  volume          = 846,
  doi             = {10.1017/jfm.2018.327},
  journal         = {Journal of Fluid Mechanics},
  author          = {Taira, Kunihiko and Nakao, Hiroya},
  year            = 2018,
  pages           = {R2}
}

@article{mauroy2013isostables,
  title           = {Isostables, isochrons, and Koopman spectrum for the
                  action–angle representation of stable fixed point dynamics},
  journal         = {Physica D: Nonlinear Phenomena},
  volume          = 261,
  pages           = {19-30},
  year            = 2013,
  issn            = {0167-2789},
  doi             = {https://doi.org/10.1016/j.physd.2013.06.004},
  author          = {A. Mauroy and I. Mezić and J. Moehlis},
}

@article{schmid2010dynamic,
  title           = {Dynamic mode decomposition of numerical and experimental
                  data},
  volume          = 656,
  DOI             = {10.1017/S0022112010001217},
  journal         = {Journal of Fluid Mechanics},
  author          = {Schmid, Peter J.},
  year            = 2010,
  pages           = {5–28}
}

@book{kutz2016dynamic,
  title           = {Dynamic mode decomposition: data-driven modeling of complex
                  systems},
  author          = {Kutz, J Nathan and Brunton, Steven L and Brunton, Bingni W
                  and Proctor, Joshua L},
  year            = 2016,
  address         = {Philadelphia},
  publisher       = {SIAM},
  address         = {USA}
}

@article{williams2015data,
  author          = {Williams, Matthew O. and Kevrekidis, Ioannis G. and Rowley,
                  Clarence W.},
  title           = {A Data--Driven Approximation of the Koopman Operator:
                  Extending Dynamic Mode Decomposition},
  journal         = {Journal of Nonlinear Science},
  year            = 2015,
  month           = {Dec},
  day             = 01,
  volume          = 25,
  number          = 6,
  pages           = {1307-1346},
  issn            = {1432-1467},
  doi             = {10.1007/s00332-015-9258-5},
}

@article{lusch2018deep,
  author          = {Lusch, Bethany and Kutz, J. Nathan and Brunton, Steven L.},
  title           = {Deep learning for universal linear embeddings of nonlinear
                  dynamics},
  journal         = {Nature Communications},
  year            = 2018,
  month           = {Nov},
  day             = 23,
  volume          = 9,
  number          = 1,
  pages           = 4950,
  issn            = {2041-1723},
  doi             = {10.1038/s41467-018-07210-0},
}

@article{takeishi2017learning,
  author          = {Takeishi, Naoya and Kawahara, Yoshinobu and Yairi,
                  Takehisa},
  booktitle       = {Advances in Neural Information Processing Systems},
  editor          = {I. Guyon and U. Von Luxburg and S. Bengio and H. Wallach
                  and R. Fergus and S. Vishwanathan and R. Garnett},
  publisher       = {Curran Associates, Inc.},
  title           = {Learning Koopman Invariant Subspaces for Dynamic Mode
                  Decomposition},
  volume          = 30,
  year            = 2017
}

@article{otto2019linearly,
  author          = {Otto, Samuel E. and Rowley, Clarence W.},
  title           = {Linearly Recurrent Autoencoder Networks for Learning
                  Dynamics},
  journal         = {SIAM Journal on Applied Dynamical Systems},
  volume          = 18,
  number          = 1,
  pages           = {558-593},
  year            = 2019,
  doi             = {10.1137/18M1177846},
}

@article{namura2022estimating,
  title           = {Estimating asymptotic phase and amplitude functions of
                  limit-cycle oscillators from time series data},
  author          = {Namura, Norihisa and Takata, Shohei and Yamaguchi,
                  Katsunori and Kobayashi, Ryota and Nakao, Hiroya},
  journal         = {Phys. Rev. E},
  volume          = 106,
  issue           = 1,
  pages           = 014204,
  numpages        = 12,
  year            = 2022,
  month           = {Jul},
  publisher       = {American Physical Society},
  doi             = {10.1103/PhysRevE.106.014204},
}

@article{yawata2024phase,
  author          = {Yawata, Koichiro and Fukami, Kai and Taira, Kunihiko and
                  Nakao, Hiroya},
  title           = {Phase autoencoder for limit-cycle oscillators},
  journal         = {Chaos: An Interdisciplinary Journal of Nonlinear Science},
  volume          = 34,
  number          = 6,
  pages           = 063111,
  year            = 2024,
  month           = 06,
  issn            = {1054-1500},
  doi             = {10.1063/5.0205718},
}

@article{raissi2019physics,
  author          = {M. Raissi and P. Perdikaris and G.E. Karniadakis},
  title           = {Physics-informed neural networks: A deep learning framework
                  for solving forward and inverse problems involving nonlinear
                  partial differential equations},
  journal         = {Journal of Computational Physics},
  volume          = 378,
  pages           = {686-707},
  year            = 2019,
  issn            = {0021-9991},
  doi             = {https://doi.org/10.1016/j.jcp.2018.10.045},
}

@article{karniadakis2021physics,
  author          = {Karniadakis, George Em and Kevrekidis, Ioannis G. and Lu,
                  Lu and Perdikaris, Paris and Wang, Sifan and Yang, Liu},
  title           = {Physics-informed machine learning},
  journal         = {Nature Reviews Physics},
  year            = 2021,
  month           = {Jun},
  day             = 01,
  volume          = 3,
  number          = 6,
  pages           = {422-440},
  issn            = {2522-5820},
  doi             = {10.1038/s42254-021-00314-5},
}

@article{wilson2016isostable,
  title           = {Isostable reduction of periodic orbits},
  author          = {Wilson, Dan and Moehlis, Jeff},
  journal         = {Phys. Rev. E},
  volume          = 94,
  issue           = 5,
  pages           = 052213,
  numpages        = 7,
  year            = 2016,
  month           = {Nov},
  publisher       = {American Physical Society},
  doi             = {10.1103/PhysRevE.94.052213},
}

@article{mauroy2016global,
  author          = {Mauroy, Alexandre and Mezić, Igor},
  journal         = {IEEE Transactions on Automatic Control},
  title           = {Global Stability Analysis Using the Eigenfunctions of the
                  Koopman Operator},
  year            = 2016,
  volume          = 61,
  number          = 11,
  pages           = {3356-3369},
  doi             = {10.1109/TAC.2016.2518918}
}

@article{rossler1976equation,
  title           = {An equation for continuous chaos},
  journal         = {Physics Letters A},
  volume          = 57,
  number          = 5,
  pages           = {397-398},
  year            = 1976,
  issn            = {0375-9601},
  doi             = {10.1016/0375-9601(76)90101-8},
  author          = {O.E. Rössler},
}

@article{fitzhugh1961impulses,
  author          = {FitzHugh, Richard},
  title           = {Impulses and Physiological States in Theoretical Models of
                  Nerve Membrane},
  journal         = {Biophysical Journal},
  year            = 1961,
  month           = {Jul},
  day             = 01,
  publisher       = {Elsevier},
  volume          = 1,
  number          = 6,
  pages           = {445-466},
  issn            = {0006-3495},
  doi             = {10.1016/S0006-3495(61)86902-6},
}

@article{kingma2014adam,
  title           = {Adam: A method for stochastic optimization},
  author          = {Kingma, Diederik P and Ba, Jimmy},
  journal         = {arXiv preprint arXiv:1412.6980},
  year            = 2014,
  doi             = {10.48550/arXiv.1412.6980 Focus to learn more }
}

@article{liu1989limited,
  author          = {Liu, Dong C. and Nocedal, Jorge},
  title           = {On the limited memory BFGS method for large scale
                  optimization},
  journal         = {Mathematical Programming},
  year            = 1989,
  month           = {Aug},
  day             = 01,
  volume          = 45,
  number          = 1,
  pages           = {503-528},
  issn            = {1436-4646},
  doi             = {10.1007/BF01589116},
}

@article{paszke2019pytorch,
  author          = {Paszke, Adam and Gross, Sam and Massa, Francisco and Lerer,
                  Adam and Bradbury, James and Chanan, Gregory and Killeen,
                  Trevor and Lin, Zeming and Gimelshein, Natalia and Antiga,
                  Luca and Desmaison, Alban and Kopf, Andreas and Yang, Edward
                  and DeVito, Zachary and Raison, Martin and Tejani, Alykhan and
                  Chilamkurthy, Sasank and Steiner, Benoit and Fang, Lu and Bai,
                  Junjie and Chintala, Soumith},
  booktitle       = {Advances in Neural Information Processing Systems},
  editor          = {H. Wallach and H. Larochelle and A. Beygelzimer and F.
                  d\textquotesingle Alch\'{e}-Buc and E. Fox and R. Garnett},
  publisher       = {Curran Associates, Inc.},
  title           = {PyTorch: An Imperative Style, High-Performance Deep
                  Learning Library},
  volume          = 32,
  year            = 2019
}

@article{mircheski2023phase,
  author          = {Mircheski, Petar and Zhu, Jinjie and Nakao, Hiroya},
  title           = {Phase-amplitude reduction and optimal phase locking of
                  collectively oscillating networks},
  journal         = {Chaos: An Interdisciplinary Journal of Nonlinear Science},
  volume          = 33,
  number          = 10,
  pages           = 103111,
  year            = 2023,
  month           = 10,
  issn            = {1054-1500},
  doi             = {10.1063/5.0161119},
}

@article{ermentrout1996type,
  author          = {Ermentrout, Bard},
  title           = {Type I Membranes, Phase Resetting Curves, and Synchrony},
  journal         = {Neural Computation},
  volume          = 8,
  number          = 5,
  pages           = {979-1001},
  year            = 1996,
  month           = 07,
  issn            = {0899-7667},
  doi             = {10.1162/neco.1996.8.5.979},
}

@article{chen2018neural,
  author          = {Chen, Ricky T. Q. and Rubanova, Yulia and Bettencourt,
                  Jesse and Duvenaud, David K},
  booktitle       = {Advances in Neural Information Processing Systems},
  editor          = {S. Bengio and H. Wallach and H. Larochelle and K. Grauman
                  and N. Cesa-Bianchi and R. Garnett},
  publisher       = {Curran Associates, Inc.},
  title           = {Neural Ordinary Differential Equations},
  volume          = 31,
  year            = 2018
}

@article{nagumo1962,
  title           = {An Active Pulse Transmission Line Simulating Nerve Axon},
  volume          = 50,
  ISSN            = {0096-8390},
  DOI             = {10.1109/jrproc.1962.288235},
  number          = 10,
  journal         = {Proceedings of the IRE},
  publisher       = {Institute of Electrical and Electronics Engineers (IEEE)},
  author          = {Nagumo, J. and Arimoto, S. and Yoshizawa, S.},
  year            = 1962,
  month           = Oct,
  pages           = {2061–2070}
}

@article{brunton2016discovering,
  author          = {Steven L. Brunton and Joshua L. Proctor and J. Nathan Kutz
                  },
  title           = {Discovering governing equations from data by sparse
                  identification of nonlinear dynamical systems},
  journal         = {Proceedings of the National Academy of Sciences},
  volume          = 113,
  number          = 15,
  pages           = {3932-3937},
  year            = 2016,
  doi             = {10.1073/pnas.1517384113},
}

@article{brown2004phase,
  author          = {Brown, Eric and Moehlis, Jeff and Holmes, Philip},
  journal         = {Neural Computation},
  title           = {On the Phase Reduction and Response Dynamics of Neural
                  Oscillator Populations},
  year            = 2004,
  volume          = 16,
  number          = 4,
  pages           = {673-715},
  doi             = {10.1162/089976604322860668}
}

@article{mauroy2012use,
  author          = {Mauroy, A. and Mezić, I.},
  title           = {On the use of Fourier averages to compute the global
                  isochrons of (quasi)periodic dynamics},
  journal         = {Chaos: An Interdisciplinary Journal of Nonlinear Science},
  volume          = 22,
  number          = 3,
  pages           = 033112,
  year            = 2012,
  month           = 07,
  issn            = {1054-1500},
  doi             = {10.1063/1.4736859},
}

@article{yamamoto2025gaussian,
  title           = {Gaussian Process Phase Interpolation for estimating the
                  asymptotic phase of a limit cycle oscillator from time series
                  data},
  journal         = {Chaos, Solitons \& Fractals},
  volume          = 191,
  pages           = 115913,
  year            = 2025,
  issn            = {0960-0779},
  doi             = {10.1016/j.chaos.2024.115913},
  author          = {Taichi Yamamoto and Hiroya Nakao and Ryota Kobayashi},
}

@article{namura2023designing,
  author          = {Namura, Norihisa and Ishii, Tsubasa and Nakao, Hiroya},
  journal         = {IEEE Transactions on Automatic Control},
  title           = {Designing 2-D Limit-Cycle Oscillators With Prescribed
                  Trajectories and Phase-Response Characteristics},
  year            = 2024,
  volume          = 69,
  number          = 5,
  pages           = {3144-3157},
  doi             = {10.1109/TAC.2023.3337728}
}

@article{osinga2010continuation,
  author          = {Osinga, Hinke M. and Moehlis, Jeff},
  title           = {Continuation-based Computation of Global Isochrons},
  journal         = {SIAM Journal on Applied Dynamical Systems},
  volume          = 9,
  number          = 4,
  pages           = {1201-1228},
  year            = 2010,
  doi             = {10.1137/090777244},
}

@article{perez2020global,
  author          = {Pérez-Cervera, Alberto and M-Seara, Tere and Huguet, Gemma},
  title           = {Global phase-amplitude description of oscillatory dynamics
                  via the parameterization method},
  journal         = {Chaos: An Interdisciplinary Journal of Nonlinear Science},
  volume          = 30,
  number          = 8,
  pages           = 083117,
  year            = 2020,
  month           = 08,
  issn            = {1054-1500},
  doi             = {10.1063/5.0010149},
}

@article{nakao2005synchrony,
  title           = {Synchrony of limit-cycle oscillators induced by random
                  external impulses},
  author          = {Nakao, Hiroya and Arai, Ken-suke and Nagai, Ken and Tsubo,
                  Yasuhiro and Kuramoto, Yoshiki},
  journal         = {Phys. Rev. E},
  volume          = 72,
  issue           = 2,
  pages           = 026220,
  numpages        = 13,
  year            = 2005,
  month           = {Aug},
  publisher       = {American Physical Society},
  doi             = {10.1103/PhysRevE.72.026220},
}

@article{adler1946,
  author          = {Adler, R.},
  journal         = {Proceedings of the IRE},
  title           = {A Study of Locking Phenomena in Oscillators},
  year            = 1946,
  volume          = 34,
  number          = 6,
  pages           = {351-357},
  doi             = {10.1109/JRPROC.1946.229930}
}

@article{kato2021optimization,
  author          = {Kato, Yuzuru and Zlotnik, Anatoly and Li, Jr-Shin and
                  Nakao, Hiroya},
  title           = {Optimization of periodic input waveforms for global
                  entrainment of weakly forced limit-cycle oscillators},
  journal         = {Nonlinear Dynamics},
  year            = 2021,
  month           = {Aug},
  day             = 01,
  volume          = 105,
  number          = 3,
  pages           = {2247-2263},
  issn            = {1573-269X},
  doi             = {10.1007/s11071-021-06703-w},
}

@article{takata2021fast,
  author          = {Takata, Shohei and Kato, Yuzuru and Nakao, Hiroya},
  title           = {Fast optimal entrainment of limit-cycle oscillators by
                  strong periodic inputs via phase-amplitude reduction and
                  Floquet theory},
  journal         = {Chaos: An Interdisciplinary Journal of Nonlinear Science},
  volume          = 31,
  number          = 9,
  pages           = 093124,
  year            = 2021,
  month           = 09,
  issn            = {1054-1500},
  doi             = {10.1063/5.0054603},
}

@article{zlotnik2013optimal,
  title           = {Optimal Waveform for Fast Entrainment of Weakly Forced
                  Nonlinear Oscillators},
  author          = {Zlotnik, Anatoly and Chen, Yifei and Kiss, Istv\'an Z. and
                  Tanaka, Hisa-Aki and Li, Jr-Shin},
  journal         = {Phys. Rev. Lett.},
  volume          = 111,
  issue           = 2,
  pages           = 024102,
  numpages        = 5,
  year            = 2013,
  month           = {Jul},
  publisher       = {American Physical Society},
  doi             = {10.1103/PhysRevLett.111.024102},
}

@article{fukami2024data,
  title           = {Data-driven transient lift attenuation for extreme vortex
                  gust–airfoil interactions},
  volume          = 992,
  DOI             = {10.1017/jfm.2024.592},
  journal         = {Journal of Fluid Mechanics},
  author          = {Fukami, Kai and Nakao, Hiroya and Taira, Kunihiko},
  year            = 2024,
  pages           = {A17}
}

@article{yawata2025phase,
  title           = {Phase autoencoder for rapid data-driven synchronization of
                  rhythmic spatiotemporal patterns},
  author          = {Yawata, Koichiro and Sakuma, Ryo and Fukami, Kai and Taira,
                  Kunihiko and Nakao, Hiroya},
  journal         = {Phys. Rev. E},
  volume          = 112,
  issue           = 6,
  pages           = 064211,
  numpages        = 13,
  year            = 2025,
  month           = {Dec},
  publisher       = {American Physical Society},
  doi             = {10.1103/yzwt-lsqt},
}

@article{moehlis2006optimal,
  author          = {Moehlis, Jeff and Shea-Brown, Eric and Rabitz, Herschel},
  title           = {Optimal Inputs for Phase Models of Spiking Neurons},
  journal         = {Journal of Computational and Nonlinear Dynamics},
  volume          = 1,
  number          = 4,
  pages           = {358-367},
  year            = 2006,
  month           = 06,
  issn            = {1555-1415},
  doi             = {10.1115/1.2338654},
}

@article{wilson2014optimal,
  author          = {Wilson, Dan and Moehlis, Jeff},
  title           = {Optimal Chaotic Desynchronization for Neural Populations},
  journal         = {SIAM Journal on Applied Dynamical Systems},
  volume          = 13,
  number          = 1,
  pages           = {276-305},
  year            = 2014,
  doi             = {10.1137/120901702},
}

@article{cai2021physics,
  author          = {Cai, Shengze and Mao, Zhiping and Wang, Zhicheng and Yin,
                  Minglang and Karniadakis, George Em},
  title           = {Physics-informed neural networks (PINNs) for fluid
                  mechanics: a review},
  journal         = {Acta Mechanica Sinica},
  year            = 2021,
  month           = {Dec},
  day             = 01,
  volume          = 37,
  number          = 12,
  pages           = {1727-1738},
  issn            = {1614-3116},
  doi             = {10.1007/s10409-021-01148-1},
}

@book{press2007numerical,
  author          = {Press, William H. and Teukolsky, Saul A. and Vetterling,
                  William T. and Flannery, Brian P.},
  title           = {Numerical Recipes 3rd Edition: The Art of Scientific
                  Computing},
  year            = 2007,
  isbn            = 0521880688,
  publisher       = {Cambridge University Press},
  address         = {USA},
  edition         = 3
}

@article{hannam2022isochron,
  author          = { Hannam, James and Krauskopf, Bernd and Osinga, Hinke M},
  title           = {Isochron foliations and global bifurcations: a case study},
  journal         = {Transactions of Mathematics and Its Applications},
  volume          = 6,
  number          = 2,
  pages           = {tnac002},
  year            = 2022,
  month           = 01,
  issn            = {2398-4945},
  doi             = {10.1093/imatrm/tnac002},
}

@article{huguet2013computation,
  author          = {Huguet, Gemma and de la Llave, Rafael},
  title           = {Computation of Limit Cycles and Their Isochrons: Fast
                  Algorithms and Their Convergence},
  journal         = {SIAM Journal on Applied Dynamical Systems},
  volume          = 12,
  number          = 4,
  pages           = {1763-1802},
  year            = 2013,
  doi             = {10.1137/120901210},
}

@article{guckenheimer1975isochrons,
  author          = {Guckenheimer, J.},
  title           = {Isochrons and phaseless sets},
  journal         = {Journal of Mathematical Biology},
  year            = 1975,
  month           = {Sep},
  day             = 01,
  volume          = 1,
  number          = 3,
  pages           = {259-273},
  issn            = {1432-1416},
  doi             = {10.1007/BF01273747},
}

@book{goodfellow2016deep,
  title           = {Deep learning},
  author          = {Goodfellow, Ian and Bengio, Yoshua and Courville, Aaron and
                  Bengio, Yoshua},
  volume          = 1,
  number          = 2,
  year            = 2016,
  address         = {Cambridge, MA},
  publisher       = {MIT press Cambridge}
}

\end{document}